\title{\bf High-order integral equation methods for problems of scattering
  by bumps and cavities on half-planes}
\author{Carlos P\'erez-Arancibia\footnote{Computing \& Mathematical Sciences, California Institute of Technology, \texttt{cperezar@caltech.edu}.}\quad and\quad  Oscar P. Bruno\footnote{Computing \& Mathematical Sciences, California Institute of Technology, \texttt{obruno@caltech.edu}.}}
\date{}
\date{\today}

\documentclass[hidelinks,11pt]{article}

\newcommand{\de}{\,\text{d}}                               
\newcommand{\e}{\operatorname{e}}                               
                              
\newcommand{\ney}{\boldsymbol{y}}                          
\newcommand{\nex}{\boldsymbol{x}}

\newcommand{\ner}{\mathbf{r}}

\newcommand{\normal}{{\mathbf{n}}}

\newcommand{\p}{\partial}

\newcommand{\real}{\mathfrak{Re}}

\newcommand{\inte}{\mathrm{int}}
\newcommand{\exte}{\mathrm{ext}}

\newcommand{\R}{\mathbb{R}}       
\newcommand{\C}{\mathbb{C}}       
  
\usepackage{multirow}
\usepackage{amsmath}
\usepackage{amsfonts}
\usepackage{amsmath}
\usepackage{amssymb}  
\usepackage{graphicx}
\usepackage{caption}
\usepackage{subcaption}
\begin{document}
\maketitle


\begin{abstract}
  This paper presents high-order integral equation methods for
  evaluation of electromagnetic wave scattering by dielectric bumps
  and dielectric cavities on perfectly conducting or dielectric
  half-planes. In detail, the algorithms introduced in this paper
  apply to eight classical scattering problems, namely: scattering by
  a dielectric bump on a perfectly conducting or a dielectric
  half-plane, and scattering by a filled, overfilled or void
  dielectric cavity on a perfectly conducting or a dielectric
  half-plane. In all cases field representations based on single-layer
  potentials for appropriately chosen Green functions are used.  The
  numerical {\em far fields and near fields} exhibit excellent
  convergence as discretizations are refined---even at and around
  points where singular fields and infinite currents exist.

\end{abstract}

\maketitle 

\section{\label{sec:introduction}Introduction}
\begin{figure}
	\begin{subfigure}[b]{0.3\textwidth}		   		
  	 	\centering	
		\includegraphics[width=\textwidth]{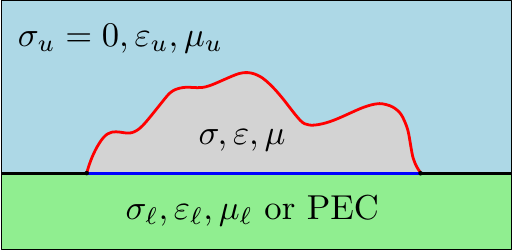}
		\caption{Dielectric bump on a half-plane.}
		\label{fig:bump}
	\end{subfigure}     \qquad
	\centering
	    \begin{subfigure}[b]{0.3\textwidth}     		
        \includegraphics[width=\textwidth]{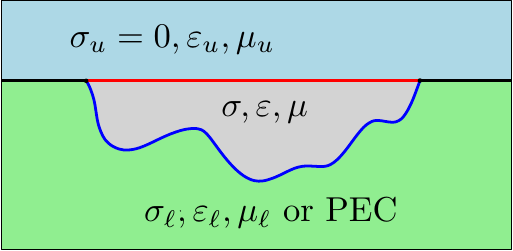}
        \caption{Dielectric-filled cavity on a half-plane.}
        \label{fig:cavity}
	\end{subfigure}\bigskip\\
			\centering
	    \begin{subfigure}[b]{0.3\textwidth}     		
        \includegraphics[width=\textwidth]{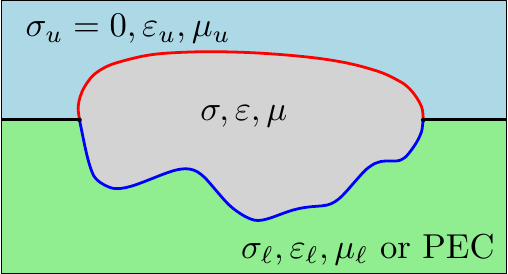}
        \caption{Dielectric-overfilled cavity on a half-plane.}
        \label{fig:overfilled-cavity}
	\end{subfigure} \qquad
	\centering
	    \begin{subfigure}[b]{0.3\textwidth}     		
        \includegraphics[width=\textwidth]{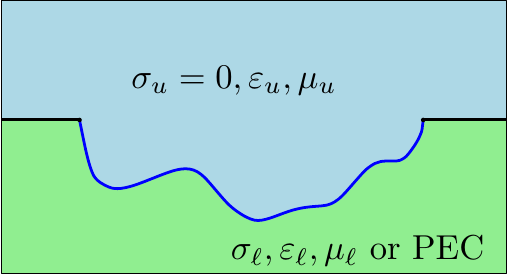}
        \caption{Void cavity on a half-plane.\newline}
        \label{fig:void-cavity}
	\end{subfigure}

    \caption{Schematics of the eight physical problems considered in this paper.}\label{fig:domain_figures}
\end{figure}

This paper presents high-order integral equation methods for the
numerical solution of problems of scattering of a plane
electromagnetic wave by cylindrical dielectric defects at the
interface between two half-planes.  Eight such classical problems are
tackled in this contribution: scattering by a dielectric bump on 1) a
perfectly electrically conducting (PEC) or 2) a dielectric half-plane
(Fig.~\ref{fig:bump}); scattering by a dielectric-filled cavity
on 3) a perfectly-conducting or 4) a dielectric half-plane
(Fig.~\ref{fig:cavity}); scattering by a dielectric-overfilled
cavity on 5) a perfectly-conducting or 6) a dielectric half-plane
(Fig.~\ref{fig:overfilled-cavity}); and scattering by a void cavity on
7) a perfectly-conducting or 8) a dielectric half-plane
(Fig.~\ref{fig:void-cavity}).  From a mathematical
perspective these eight different physical problems reduce to just
three problem types for which this paper provides numerical
solutions on the basis of highly accurate and efficient boundary
integral equation methods.

In all cases the proposed methods utilize field representations based
on single-layer potentials for appropriately chosen Green
functions. As is known, such single-layer formulations lead to
non-invertible integral equations at certain spurious
resonances---that is, for wavenumbers that coincide with interior
Dirichlet eigenvalues for a certain differential operator---either the
Laplace operator or an elliptic differential operator with piecewise
constant coefficients (see Sec.~\ref{sec:non_uniqueness} for details).
We nevertheless show that solutions {\em for all wavenumbers} can be
obtained from such non-invertible formulations---including wavenumbers
at which non-invertible integral equations result. Our method in these
regards relies on the analyticity of the PDE solution as a function of
the wavenumber together with a certain approach based on use of
Chebyshev approximation.

(The use of field representations which give rise to non-invertible
operators is advantageous in two main ways: on one hand this strategy
allows one to bypass the need to utilize hypersingular operators,
whose evaluation is computationally expensive and, otherwise, highly
challenging near corner points; and, on the other hand, it leads to
systems of integral equations containing fewer integral
operators---with associated reduced computational cost.)

The problems considered in this paper draw considerable interest in a
wide range of settings. For example, the problem of scattering by
bumps and cavities on a (perfect or imperfect) conducting half-plane
is important in the study of the radio-frequency absorption and
electric and magnetic field enhancement that arises from surface
roughness~\cite{RUPPIN:1981,ZHANG:2009}. The problem of scattering by
open groove cavities on a conducting plane, in turn, impacts on a
variety of technologies, with applicability to design of cavity-backed
antennas, non-destructive evaluation of material surfaces, and more
recently, modeling of extraordinary transmission of light and
plasmonics resonance, amongst many others (e.g.~\cite{ALAVIKIA:2011}
and references therein).

There is vast literature concerning the types of problems considered
in this paper.  For a circular bump a separation-of-variables
analytical Fourier-Bessel expansion exists~\cite{RAYLEIGH:1907}. Related 
semi-analytical separation-of-variables solutions are available for other simple
configurations, such as semi-circular cavities and rectangular bumps
and cavities (e.g.~\cite{BYUN:1998,BYUN:1998:II,EOM:1993,
LEE:2012,PARK:1993,PARK:1993:II,PARK:1993:III,TYZHNENKO:2004,
TYZHNENKO:2005,YU:2002} and references therein), while solutions based 
on Fourier-type integral representations, mode matching techniques 
and staircase approximation of the geometry are available for more 
general domains (e.g.~\cite{BASHA:2007} and references 
therein). Even for simple configurations, such as a circular cavity or bump on a perfectly conduction plane, the semi-analytical 
separation-of-variables method requires solution of an infinite 
dimensional linear system of equations that must be truncated to 
an $n\times n$ system and solved numerically \cite{HINDERS:1991,
PARK:1993,PARK:1993:III,SACHDEVA:1977,TYZHNENKO:2004,TYZHNENKO:2005}. 
As it happens, the resulting (full) matrix is extremely
ill-conditioned for large values of $n$. In practice only limited
accuracy results from use of such algorithms: use of small values of
$n$ naturally produces limited accuracy, while for large values of $n$
matrix ill-conditioning arises as an accuracy limiting element.

Finite element and finite difference methods of low order of accuracy
have been used extensively over the last
decade
\cite{ALAVIKIA:2009,ALAVIKIA:2011,BAO:2005,DU:2011,LI:2013,VAN:2003,
WANG:2008,WOOD:2006}. As
is well known, finite element and finite difference methods lead to
sparse linear systems. However, in order to satisfy the Sommerfeld
radiation condition at infinity, a relatively large computational
domain containing the scatterer must be utilized (unless a non-local
boundary condition is used, with a consequent loss of sparsity). In
view of the large required computational domains (or large coupled
systems of equations for methods that use non-local domain truncation)
and their low-order convergence (especially around corners where
fields are singular and currents are infinite), these methods yield
very slow convergence, and, therefore, for adequately accurate
solutions, they require use of large numbers of unknowns and a high
computational cost.

Boundary integral equation methods, on the other hand, lead to linear
systems of reduced dimensionality, the associated solutions
automatically satisfy the condition of radiation at infinity, and,
unlike finite element methods, they do not suffer from dispersion
errors.  Integral equation methods have been used previously for the
solution of the problem of scattering by an empty and
dielectric-filled cavity on a perfectly conducting half-plane; see
e.g.~\cite{HOWE:2003,WOOD:1999, WANG:2003}. However, previous integral
approaches for these problems are based on use of low-order numerical
algorithms and, most importantly, they do not accurately account for
singular field behavior at corners---and, thus, they may not be
sufficiently accurate for evaluation of important physical mechanisms
that arise from singular electrical currents and local fields at and
around corners.

The present paper is organized as follows.
Sec.~\ref{sec:scattering_problem} presents a brief description of the
various problems at hand and Sec.~\ref{sec:integral_formulation}
introduces a new set of integral equations for their treatment.
Sec.~\ref{sec:numerics} then describes the high-order solvers we have
developed for the numerical solution of these integral equations,
which include full resolution of singular fields at corners. The
excellent convergence properties of the equations and algorithms
introduced in this text are demonstrated in
Sec.~\ref{sec:numerical_results}. In particular, the high accuracy of
the new methods in presence of corner singularities can be used to
evaluate the effects of corner singularities on currents and local
fields on and around bumps and cavities, and, thus, on important
physical observables such as absorption, extraordinary transmission,
cavity resonance, etc.

\section{\label{sec:scattering_problem}Scattering problem}  
All the problems considered in this contribution can be described
mathematically following the compact depiction presented in
Fig.~\ref{fig:mathematical_description}. Thus, a plane wave $\bold
H^{\mathrm{inc}}(\nex) = \bold H^0\e^{i\bold k\cdot\nex}$, $\bold
E^{\mathrm{inc}}(\nex) = \bold E^0\e^{i\bold k\cdot\nex}$ with wave
vector $\bold{k}=k_3(\cos\alpha,\sin\alpha)$ impinges on a cavity
formed by the subdomains $\Omega_1$ and $\Omega_2$ which lies on the
boundary of an otherwise planar horizontal interface between the
infinite subdomains $\Omega_3$ and $\Omega_4$. As is well-known, the
$z$ components $u=E_z$ and $u=H_z$ of the total electric and magnetic
field satisfy the Helmholtz equation
\begin{equation}
\Delta u +k^2_j u=0\quad\mbox{in}\quad\Omega_j, \label{eq:helmholtz_equation}
\end{equation}
where, letting $\omega>0$, $\varepsilon_j>0$, $\mu_j>0$ and $\sigma_j\geq 0$
denote the angular frequency, the electric permittivity, the magnetic
permeability and the electrical conductivity, the wavenumber $k_j$
($\Im(k_j) >0$) is defined by $k^2_j=
\omega^2(\varepsilon_j+i\sigma_j/\omega)\mu_j$, $1\leq j\leq 4$
. Throughout this paper it is assumed that $\Omega_3$ is a lossless
medium ($\sigma_3= 0$).

In order to formulate transmission problems for the transverse components of the electromagnetic field, $u$ is expressed as 
\begin{equation}
u = \left\{\begin{array}{lcc}u_1&\mathrm{in}&\Omega_1,\\
u_2&\mathrm{in}&\Omega_2,\\
u_3+f&\mathrm{in}&\Omega_3,\\
u_4+f&\mathrm{in}&\Omega_4,\\
\end{array}\right.\label{eq:decomposition_field}
\end{equation}
where $f$ is the solution (presented below in this section) of the
problem of scattering by the lower half-plane \emph{in absence of the
  dielectric defect}.

Additionally, $u$ satisfies the  transmission conditions
\begin{equation}
\begin{array}{rclcl}
u_i-u_j&=&g,\medskip\\
\displaystyle\frac{1}{\beta_i}\frac{\p u_i}{\p \normal}-\frac{1}{\beta_j}\frac{\p u_j}{\p \normal}&=&\displaystyle\frac{1}{\beta_j}\frac{\p g}{\p\normal},
\end{array}\label{eq:transmission_conditions}
\end{equation}
at the interface $\Gamma_{ij}$ between $\Omega_i$ and $\Omega_j$,
where $\beta_j = \mu_j$ in TM-polarization and $\beta_j =
\varepsilon_j+i\sigma_j/\omega$ in TE-polarization. For each one of
the problems considered in this paper
Equations~(\ref{eq:transmission_conditions}) with $g=f$ are satisfied
on $\Gamma_{13}$. In the case in which $\Omega_4$ is filled by a
dielectric material the transmission conditions
(\ref{eq:transmission_conditions}) are also satisfied with boundary
data $g=f$ on $\Gamma_{24}$, and they are satisfied with boundary
data $g=0$ on $\Gamma_{34}$.  On the other hand, when $\Omega_4$ is a
perfectly conducting half-plane, $u_4=0$ and boundary conditions
 \begin{equation}
u_j=0\quad \mbox{and}\quad \frac{\p u_j}{\p \normal}=0,\quad j=2,3
\label{eq:PEC_conditions}
\end{equation} 
are satisfied on $\Gamma_{j4}$ in TM- and TE-polarization, respectively. Additionally, the scattering fields $u_j$, $j=3,4$  fulfill the Sommerfeld radiation condition at infinity.

The solution $f$ of the problem of scattering by the lower half-plane
in absence of the dielectric defect (which provides the necessary
source term in~\eqref{eq:decomposition_field}) can be computed
explicitly for each one of the problems considered in this paper. For
the problems in which $\Omega_4$ is a perfectly flat PEC half-plane
the total field is given by $f(\nex) = E^0_z(\e^{i\bold
  k\cdot\nex}-\e^{i\overline{\bold k}\cdot\nex})$ in TM-polarization,
and $f(\nex) = H^0_z(\e^{i\bold
  k\cdot\nex}+\e^{i\overline{\bold{k}}\cdot\nex})$ in TE-polarization,
where $\bold{k} =k_3 (\cos\alpha,\sin\alpha)$ and $\overline{\bold{k}}
=k_3 (\cos\alpha,-\sin\alpha)$. For the problems in which $\Omega_4$
is a flat dielectric half-plane, in turn, the total field is given by
\begin{equation*}
f(\nex) = \left\{\begin{array}{ccc} 
E^0_z(\e^{i\bold k\cdot\nex}+R^{\mbox{\tiny TM}}\e^{i\bar{\bold k}\cdot\nex})&\mathrm{in}&\Omega_3,\\
E^0_zT^{\mbox{\tiny TM}}\e^{i\widetilde{\bold k}\cdot\nex}&\mathrm{in}&\Omega_4,
\end{array}\right.
\end{equation*}
and
\begin{equation*}
f(\nex) = \left\{\begin{array}{ccc} 
H^0_z(\e^{i\bold k\cdot\nex}+R^{\mbox{\tiny TE}}\e^{i\bar{\bold k}\cdot\nex})&\mathrm{in}&\Omega_3,\\
H^0_zT^{\mbox{\tiny TE}}\e^{i\widetilde{\bold k}\cdot\nex}&\mathrm{in}&\Omega_4,
\end{array}\right.
\end{equation*}
in TM- and TE-polarization respectively, where 
$$
T^{\mathrm{TM,TE}} = \frac{2\beta_4k_3}{\beta_4 k_3+\lambda \beta_3k_4},\quad R^{\mathrm{TM,TE}} = \frac{\beta_4 k_3-\lambda \beta_3 k_4}{\beta_4 k_3+\lambda \beta_3 k_4},\quad\lambda = \frac{\sqrt{1-k_3^2/k_4^2\cos^2(\alpha)}}{|\sin(\alpha)|}
$$ and
$\widetilde{\bold k} =k_4
\left(k_3/k_4\cos(\alpha),-\sqrt{1-k_3^2/k_4^2\cos^2(\alpha)}\right)$
(using the square root function $\sqrt{z}$ determined by the relation
$-\pi < \arg(z)\leq \pi$---so that, in particular, $\sqrt{-1} = i$).

\section{\label{sec:integral_formulation} Integral equation formulations}
\begin{figure}[h!]
\centering	
\includegraphics[width=7.5cm]{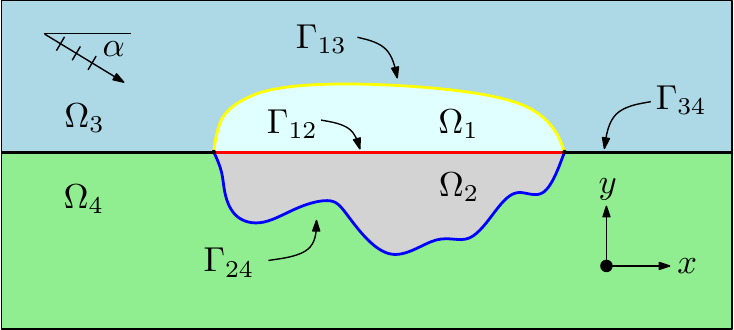}
\caption{Compact mathematical description of the problems considered in this paper.}\label{fig:mathematical_description}
\end{figure}

Three main problem types can be identified in connection with
Fig.~\ref{fig:mathematical_description}, namely \emph{Problem
  Type~I}, where transmission conditions
\eqref{eq:transmission_conditions} are imposed on $\Gamma_{13}$ and
$\Gamma_{24}$ (which, in our context, characterize the problem of
scattering by a dielectric bump on a dielectric half-plane as well as
the problems of scattering by a filled, overfilled or empty cavity on
a dielectric half-plane); \emph{Problem Type~II}, where transmission
conditions \eqref{eq:transmission_conditions} are imposed on
$\Gamma_{13}$ and PEC boundary condition \eqref{eq:PEC_conditions} is
imposed on $\Gamma_{24}$, which applies to the problem of scattering
by a (filled, overfilled or empty) cavity on a PEC half-plane; and
\emph{Problem Type~III}, where transmission conditions
\eqref{eq:transmission_conditions} are only imposed on $\Gamma_{13}$,
with application to the problem of scattering by a dielectric bump on
a perfectly conducting half-plane. In the following three sections we
derive systems of boundary integral equations for each one of these
problem types.

\subsection{\label{sec:integral_formulation_I}Problem Type~I}
In Problem Type~I the domains $\Omega_j$ ($1\leq j\leq 4$) contain
dielectric media of finite or zero conductivity; we denote by $k_j$
the (real or complex) wavenumber in the domain $\Omega_j$.  Note that
\begin{itemize}
\item[--] For the problem of scattering by a \emph{dielectric-filled
    cavity on dielectric half-plane} we have $k_3=k_1,k_1\neq
  k_2,k_2\neq k_4$;
\item[--] For the problem of scattering by an \emph{overfilled cavity
    on dielectric half-plane} we have $k_3\neq k_1,k_1=k_2,k_2\neq
  k_4$; and
\item[--] For the problem of scattering by a \emph{void cavity on a
    dielectric half-plane} we have $k_3=k_1,k_1=k_2,k_2\neq k_4$.
\end{itemize}
To tackle the Type~I problem we express the total field $u$ by means the
single-layer-potential representation
\begin{equation}
  u=\left\{\begin{array}{lll} \mathcal S_\inte[\psi_\inte]&\mbox{in}&\Omega_1\cup\Omega_2,\medskip\\
      \mathcal S_\exte[\psi_\exte]+f&\mbox{in}&\Omega_3\cup\Omega_4,
\end{array}\right.
 \label{eq:integral_representation}
\end{equation}
in terms of the unknown density functions $\psi_\inte$ and
$\psi_\exte$ where, letting $G_{k_j}^{k_i}$ denote the Green function
of the Helmholtz equation for the two-layer medium with wavenumbers
$k_i$ and $k_j$ in the upper and lower half-planes respectively (see
Appendix~\ref{app:greens_function}), we have set
\begin{subequations}
\begin{eqnarray}
 \mathcal S_\inte[\psi](\nex) &=& \int_{\Gamma_{13}\cup\Gamma_{24}} G^{k_1}_{k_2}(\nex,\ney)\psi(\ney)\de s_{\ney},\medskip\\
\mathcal S_\exte[\psi](\nex) &=& \int_{\Gamma_{13}\cup\Gamma_{24}}G_{k_4}^{k_3}(\nex,\ney)\psi(\ney)\de s_{\ney}.
\end{eqnarray}\label{eq:single_layer_representation_I}\end{subequations}
The Green functions $G^{k_i}_{k_j}$ satisfy the transmission
conditions \eqref{eq:transmission_conditions} on $\Gamma_{ij}$ (with
$(ij)$ equal to either $(12)$ or $(34)$) and, therefore, they depend
on the polarization (through the parameters $\beta_i$ and
$\beta_j$). Note, further, that for $k_i = k_j = k$ the Green function
$G^{k_i}_{k_j}$ equals the free space Green function with wavenumber
$k$.

It is easy to check that the
representation~\eqref{eq:integral_representation} for the solution $u$
satisfies the Helmholtz equation with wavenumber $k_j$ in the domain
$\Omega_j$ ($1\leq j\leq 4$) as well as the radiation conditions at
infinity. Since the two-layer Green functions satisfy the relevant
transmission conditions on $\Gamma_{12}$ and~$\Gamma_{34}$, there
remain only two boundary conditions to be satisfied, namely, the
transmission conditions~\eqref{eq:transmission_conditions} on the
boundary of the defect $\Omega_1\cup\Omega_2$.  Using classical jump
relations~\cite{COLTON:2012} for various layer potentials, these
conditions lead to the system
\begin{equation}
\begin{array}{rrccl}
\displaystyle S^{\Gamma_{13}}_\inte[\psi_\inte] -S^{\Gamma_{13}}_{\exte}[\psi_\exte] &=& f,\medskip\\
\displaystyle \frac{\beta_3}{\beta_1}\left\{\frac{\psi_\inte}{2}-K^{\Gamma_{13}}_\inte[\psi_\inte]\right\} +\frac{\psi_\exte}{2} - K^{\Gamma_{13}}_\exte[\psi_\exte] &=& \displaystyle \frac{\p f}{\p \normal},\medskip\\
\displaystyle S^{\Gamma_{24}}_\inte[\psi_\inte] -S^{\Gamma_{24}}_{\exte}[\psi_\exte] &=& f,\medskip\\
\displaystyle \frac{\beta_4}{\beta_2}\left\{\frac{\psi_\inte}{2}-K^{\Gamma_{24}}_\inte[\psi_\inte]\right\} +\frac{\psi_\exte}{2} - K^{\Gamma_{24}}_\exte[\psi_\exte] &=& \displaystyle \frac{\p f}{\p \normal}
\end{array}\label{eq:integral_equations_I}
\end{equation}
of boundary integral equations on the open curves $\Gamma_{13}$ and
$\Gamma_{24}$ for the unknowns $\psi_\inte$ and $\psi_\exte$.  The
boundary integral operators in~\eqref{eq:integral_equations_I} for
$(ij)=(13)$ and $(ij)=(24)$ are given by
\begin{equation}
\begin{array}{l}
 S^{\Gamma_{ij}}_\inte[\psi](\nex) =\displaystyle \int_{\Gamma_{13}\cup\Gamma_{24}} G^{k_1}_{k_2}(\nex,\ney)\psi(\ney)\de s_{\ney},\ \nex\in\Gamma_{ij},
\medskip\\
 S^{\Gamma_{ij}}_\exte[\psi](\nex) =\displaystyle \int_{\Gamma_{13}\cup\Gamma_{24}}G_{k_4}^{k_3}(\nex,\ney)\psi(\ney)\de s_{\ney},\ \nex\in\Gamma_{ij},\medskip\\
 K^{\Gamma_{ij}}_\inte[\psi](\nex) =\displaystyle \int_{\Gamma_{13}\cup\Gamma_{24}} \frac{\p G^{k_1}_{k_2}}{\p\normal_{\nex}}(\nex,\ney)\psi(\ney)\de s_{\ney},\ \nex\in\Gamma_{ij},
\medskip\\
 K^{\Gamma_{ij}}_\exte[\psi](\nex) =\displaystyle \int_{\Gamma_{13}\cup\Gamma_{24}}\frac{\p G_{k_4}^{k_3}}{\p \normal_{\nex}}(\nex,\ney)\psi(\ney)\de s_{\ney},\ \nex\in\Gamma_{ij}.
\end{array}\label{eq:integral_operators_I}
\end{equation}


\subsection{\label{sec:integral_formulation_II}Problem Type II}

In Problem Type~II the domain $\Omega_4$ contains a PEC medium, and
the domains $\Omega_j$ ($1\leq j\leq 3$) contain dielectric media of
finite or zero conductivity; we denote by $k_j$ the (real or complex)
wavenumber in the domain~$\Omega_j$~($1\leq j\leq 3$).  Clearly,
\begin{itemize}
\item[--] For the problem of scattering by a \emph{dielectric-filled
    cavity on PEC half-plane} we have $k_3=k_1,k_1\neq k_2$;
\item[--] For the problem of scattering by an \emph{overfilled cavity
    on PEC half-plane} we have $k_3\neq k_1,k_1=k_2$; and
\item[--] For the problem of scattering by a \emph{void cavity on PEC
    half-plane} we have $k_3=k_1,k_1=k_2$.
\end{itemize}
For Type~II problems we express the total field $u$ by means of the
single-layer-potential representation
\begin{equation}
u=\left\{\begin{array}{lll} \mathcal S_\inte[\psi_\inte]&\mbox{in}&\Omega_1\cup\Omega_2,\medskip\\
\mathcal S_\exte[\psi_\exte]+f&\mbox{in}&\Omega_3,\medskip\\
0&\mbox{in}&\Omega_4,
\end{array}\right.
 \label{eq:integral_representation_II}
\end{equation}
where, defining $G^{k_1}_{k_2}$ as in
Sec.~\ref{sec:integral_formulation_I} and letting
$G^{k_3}_{\infty}$ denote the Green function that satisfies the PEC
boundary condition~\eqref{eq:PEC_conditions} on $\Gamma_{34}$, the
potentials above are defined by
\begin{subequations}
\begin{eqnarray}
 \mathcal S_\inte[\psi](\nex) &=& \int_{\Gamma_{13}\cup\Gamma_{24}} G^{k_1}_{k_2}(\nex,\ney)\psi(\ney)\de s_{\ney},\medskip\\
\mathcal S_\exte[\psi](\nex) &=& \int_{\Gamma_{13}}G^{k_3}_{\infty}(\nex,\ney)\psi(\ney)\de s_{\ney}.
\end{eqnarray} \label{eq:single_layer_representation_II}\end{subequations}
As mentioned in Sec.~\ref{sec:integral_formulation_I} the Green
function $G^{k_i}_{k_j}$ depends on the polarization; the same is of
course true for $G^{k_3}_{\infty}$, which is given by $
G^{k}_{\infty}(\nex,\ney)= G_k(\nex,\ney)-G_k(\bar\nex,\ney)$ in
TM-polarization, and
$G_{\infty}^{k}(\nex,\ney)=G_k(\nex,\ney)+G_k(\bar\nex,\ney)$ in
TE-polarization, where $\bar\nex=(x_1,-x_2)$ and where
$G_k(\nex,\ney)=iH_0^{(1)}(k|\nex-\ney|)/4$ is the free-space Green
function. By virtue of the integral representation~\eqref{eq:integral_representation_II} the field satisfies the Helmholtz equation in the domain $\Omega_j$ with wavenumber $k_j$ ($1 \leq j\leq 3$), the radiation condition at infinity, transmission conditions on $\Gamma_{12}$ and the PEC boundary conditions on $\Gamma_{24}$. Imposing the remaining  transmission conditions~\eqref{eq:transmission_conditions} on $\Gamma_{13}$ and  PEC boundary condition~\eqref{eq:PEC_conditions} of $\Gamma_{24}$, we obtain the equations
\begin{subequations}
\begin{equation}\begin{array}{rcl}
\displaystyle S^{\Gamma_{13}}_\inte[\psi_\inte] -S^{\Gamma_{13}}_{\exte}[\psi_\exte] &=& f,\medskip\\
\displaystyle \frac{\beta_3}{\beta_1}\left\{\frac{\psi_\inte}{2}-K^{\Gamma_{13}}_\inte[\psi_\inte]\right\} +\frac{\psi_\exte}{2} - K^{\Gamma_{13}}_\exte[\psi_\exte] &=& \displaystyle \frac{\p f}{\p \normal},
\end{array}\end{equation} on $\Gamma_{13}$ (valid for both TE and TM polarizations provided the corresponding constants~$\beta_j$ and Green functions are used)   and
\begin{eqnarray}
 \frac{\psi_\inte}{2} + K^{\Gamma_{24}}_{\inte}[\psi_\inte] =0\quad (\mbox{TE polarization})\label{eq:cavity_PEC_TE}\medskip\\
 S^{\Gamma_{24}}_\inte[\psi_\inte]  = 0\quad(\mbox{TM polarization})\label{eq:cavity_PEC_TM}
\end{eqnarray} \label{eq:integral_equations_II}\end{subequations} 
on $\Gamma_{24}$. In accordance with the definition of the single-layer potentials~\eqref{eq:single_layer_representation_II}, the boundary integral operators in~\eqref{eq:integral_equations_II} for $(ij)=(13)$ and $(ij)=(24)$ are given by 
\begin{equation}
\begin{array}{l}
 S^{\Gamma_{ij}}_\inte[\psi](\nex) = \displaystyle\int_{\Gamma_{13}\cup\Gamma_{24}} G^{k_1}_{k_2}(\nex,\ney)\psi(\ney)\de s_{\ney},\ \nex\in\Gamma_{ij},
\medskip\\
 S^{\Gamma_{ij}}_\exte[\psi](\nex) = \displaystyle\int_{\Gamma_{13}}G^{k_3}_{\infty}(\nex,\ney)\psi(\ney)\de s_{\ney},\ \nex\in\Gamma_{ij},\medskip\\
 K^{\Gamma_{ij}}_\inte[\psi](\nex) = \displaystyle\int_{\Gamma_{13}\cup\Gamma_{24}} \frac{\p G^{k_1}_{k_2}}{\p\normal_{\nex}}(\nex,\ney)\psi(\ney)\de s_{\ney},\ \nex\in\Gamma_{ij},
\medskip\\
 K^{\Gamma_{ij}}_\exte[\psi](\nex) = \displaystyle\int_{\Gamma_{13}}\frac{\p G_{\infty}^{k_3}}{\p \normal_{\nex}}(\nex,\ney)\psi(\ney)\de s_{\ney},\ \nex\in\Gamma_{ij}.
\end{array}\label{eq:integral_operators_II}
\end{equation}

\subsection{\label{sec:integral_formulation_III} Problem Type III} 
For Problem Type III the domains $\Omega_j$ ($j=1,3$) contain
dielectric media of finite or zero conductivity (the corresponding,
possibly complex, wavenumbers are denoted by $k_1$ and $k_3$), and the
domains $\Omega_j$, $j=2,4,$ contain a PEC medium. Note that
\begin{itemize}
\item[--]  For the problem of scattering by a
\emph{dielectric bump on PEC half-plane} we have~$k_3\neq k_1$.
\end{itemize}

As in the previous cases, for Type III problems the total field $u$ is expressed by means of the single-layer-potential representation
\begin{equation}
u=\left\{\begin{array}{lll} \mathcal S_\inte[\psi_\inte]&\mbox{in}&\Omega_1,\medskip\\
\mathcal S_\exte[\psi_\exte]+f&\mbox{in}&\Omega_3,\medskip\\
0&\mbox{in}&\Omega_2\cup\Omega_4,
\end{array}\right.
 \label{eq:integral_representation_III}
\end{equation}
where the potentials above are defined by
\begin{subequations}
\begin{eqnarray}
 \mathcal S_\inte[\psi](\nex) &=& \int_{\Gamma_{13}} G^{k_1}_{\infty}(\nex,\ney)\psi(\ney)\de s_{\ney},\medskip\\
 \mathcal S_\exte[\psi](\nex) &=& \int_{\Gamma_{13}} G^{k_3}_{\infty}(\nex,\ney)\psi(\ney)\de s_{\ney}.
\end{eqnarray}\label{eq:single_layer_representation_III}\end{subequations}
As mentioned in Sec.~\ref{sec:integral_formulation_II}, the Green
functions $G_\infty^{k_1}$ and~$G_\infty^{k_3}$ depend on the
polarization and satisfy the PEC boundary condition on $\Gamma_{12}$
and $\Gamma_{34}$ respectively. The total field, as given by the
potentials~\eqref{eq:single_layer_representation_III}, satisfies
Helmholtz equations with wavenumber $k_j$ in the domain $\Omega_j$,
$j=1,3$, PEC boundary condition on $\Gamma_{24}$ and $\Gamma_{12}$, as
well as the radiation condition at infinity. Imposing the transmission
conditions~\eqref{eq:transmission_conditions} on $\Gamma_{13}$ the
following system of boundary integral equations is obtained for the
unknown density functions $\psi_\inte$ and $\psi_\exte$:
\begin{equation}\begin{array}{rcl}
\displaystyle S^{\Gamma_{13}}_\inte[\psi_\inte] -S^{\Gamma_{13}}_{\exte}[\psi_\exte] &=& f,\medskip\\
\displaystyle \frac{\beta_3}{\beta_1}\left\{\frac{\psi_\inte}{2}-K^{\Gamma_{13}}_\inte[\psi_\inte]\right\} +\frac{\psi_\exte}{2} - K^{\Gamma_{13}}_\exte[\psi_\exte] &=& \displaystyle \frac{\p f}{\p \normal},\label{eq:integral_equations_III}\end{array}\end{equation} on $\Gamma_{13}$, where
the boundary integral operators are defined by 
\begin{equation}\begin{array}{l}
 S^{\Gamma_{13}}_\inte[\psi](\nex) = \displaystyle\int_{\Gamma_{13}} G^{k_1}_{\infty}(\nex,\ney)\psi(\ney)\de s_{\ney},\ \nex\in\Gamma_{13},
\medskip\\
 S^{\Gamma_{13}}_\exte[\psi](\nex) =\displaystyle \int_{\Gamma_{13}}G^{k_3}_{\infty}(\nex,\ney)\psi(\ney)\de s_{\ney},\ \nex\in\Gamma_{13},\medskip\\
 K^{\Gamma_{13}}_\inte[\psi](\nex) =\displaystyle \int_{\Gamma_{13}} \frac{\p G^{k_1}_{\infty}}{\p\normal_{\nex}}(\nex,\ney)\psi(\ney)\de s_{\ney},\ \nex\in\Gamma_{13},
\medskip\\
 K^{\Gamma_{13}}_\exte[\psi](\nex) =\displaystyle \int_{\Gamma_{13}}\frac{\p G_{\infty}^{k_3}}{\p \normal_{\nex}}(\nex,\ney)\psi(\ney)\de s_{\ney},\ \nex\in\Gamma_{13}.
\end{array}\label{eq:integral_operators_III}
\end{equation}
\section{\label{sec:numerics}Numerical method}
\subsection{Discretization of integral equations}
The integral
equations~\eqref{eq:integral_equations_I},~\eqref{eq:integral_equations_II}
and~\eqref{eq:integral_equations_III} involve either a)~Integrals over
$\Gamma_{13}\cup\Gamma_{24}$ with equality enforced on
$\Gamma_{13}\cup\Gamma_{24}$, or given by b)~Integrals over
$\Gamma_{13}$ with equality enforced on $\Gamma_{13}$. All of these
integral equations can be expressed in terms of parametrizations of
the curves $\Gamma_{13}$ and $\Gamma_{24}$, or, more precisely, in
terms of integrals of the form
\begin{eqnarray}
\int_{0}^{2\pi}L(t,\tau)\phi(\tau)\de \tau&\quad \mbox{and}\quad &\int_{0}^{2\pi}M(t,\tau)\phi(\tau)\de \tau, \label{eq:parametrized_operators}
\end{eqnarray}
with kernels
\begin{equation}\begin{array}{rcl}
   L(t,\tau) &=&  \displaystyle G(\nex(t),\ney(\tau))|\ney'(\tau)|,\medskip\\
   M(t,\tau) &=&  \displaystyle\nabla_{\nex} [G](\nex(t),\ney(\tau))\cdot \normal(t)|\ney'(\tau)|
\end{array}\label{eq:integral_kernels}
\end{equation}
where {\em i)}~Each of the functions $\nex(t)$ and $\ney(\tau)$ denote
either a parametrization for the curve $\Gamma_{13}$ or of the curve
$\Gamma_{24}$ with parameters $t$ and $\tau$ in the interval
$(0,2\pi)$; {\em ii)}~$\normal(t) = (x'_2(t),-x'_1(t))/|\nex'(t)|$
denotes the unit normal on $\Gamma_{13}$ or $\Gamma_{24}$, as
appropriate, which points outward from the defect; {\em
  iii)}~$\phi(\tau)= \psi(\ney(\tau))$, where $\psi$ stands for the
unknown density function under consideration; and {\em iv)}~$G$
denotes the relevant Green function.  Indeed, in case a)~above, the
integral over $\Gamma_{13}\cup\Gamma_{24}$ can be expressed as a sum
of integrals on $\Gamma_{13}$ and $\Gamma_{24}$. In case~b), in particular,
we  take $\nex=\ney$.

Our discretization of the integral
equations~\eqref{eq:integral_equations_I},~\eqref{eq:integral_equations_II}
and~\eqref{eq:integral_equations_III} is based on corresponding
discretizations of the
integrals~\eqref{eq:parametrized_operators}. Following~\cite{COLTON:2012}
we thus proceed by expressing the kernels~\eqref{eq:integral_kernels}
in the form
\begin{subequations}
\begin{eqnarray}
L(t, \tau)&=& L_1(t, \tau)\log r^2(t,\tau) +L_2(t, \tau),\medskip\\
M(t, \tau)&=& M_1(t, \tau)\log r^2(t,\tau) +M_2(t, \tau),\
\end{eqnarray} \label{eq:kernels}\end{subequations}
where $L_j$ and $M_j$ ($j=1,2$) are smooth functions on $(0,
2\pi)\times (0, 2\pi)$ and where $\ner(t,\tau) = \nex(t)-\ney(\tau)$
and $r(t,\tau) = |\ner(t,\tau)|$.  In cases for which $\nex(t)$ and
$\ney(\tau)$ parametrize the same open curve we have
\begin{eqnarray*}
L_1(t,\tau)  &=&-\frac{1}{4\pi}J_0(kr(t,\tau))|\ney'(\tau)|, \medskip\\
L_2(t,\tau) &=& L(t,\tau)-L_1(t, \tau)\log r^2(t,\tau),\medskip\\
M_1(t,\tau) &=& \frac{k}{4\pi}J_1(kr(t,\tau))\normal(t)\cdot \frac{\ner(t,\tau)}{r}|\ney'(\tau)|,\medskip\\
M_2(t,\tau) &=&M(t,\tau)-M_1(t, \tau)\log r^2(t,\tau).
\end{eqnarray*}	
The diagonal terms $L_2(t,t)$ and $M_2(t,t)$ can be computed exactly
by taking the limit of $L_2(t,\tau)$ and $M_2(t,\tau)$ as
$\tau\rightarrow t$ (see~\cite[p. 77]{COLTON:2012}
for details).  On the other hand, when $\nex(t)$ and $\ney(\tau)$
parametrize different curves, $L$ and $M$ are smooth on
$(0,2\pi)\times(0,2\pi)$ and, thus, $L_1=0$, $L=L_2$, $M_1=0$ and
$M=M_2$. (Note that although in the latter case $L$ and $M$ are smooth
functions, these functions are in fact {\em nearly singular}, for $t$
near the endpoints of the parameter interval $(0,2\pi)$ for the curve
$\nex$, and for $\tau$ around the corresponding endpoint of the
parameter interval for the curve $\ney$.)

Letting $K$ denote one of the integral kernels $L$ or $M$ in
equation~\eqref{eq:kernels}, in view of the discussion above $K$ may
be expressed in the form $K(t, \tau)= K_1(t, \tau)\log r^2(t,\tau)
+K_2(t, \tau)$ for smooth kernels $K_1$ and $K_2$. For a fixed~$t$
then, there are two types of integrands for which high-order
quadratures must be provided, namely integrands that are smooth in
$(0, 2\pi)$ but have singularities at the endpoints of the interval
(that arise from corresponding singularities of the densities $\phi$
at the endpoints of the open curves;
cf.~\cite{meixner:1972,van:1985,BRUNO:2009}), and integrands that
additionally have a logarithmic singularity at $\tau = t$. To handle
both singular integration problems we
follow~\cite{KRESS:1990,COLTON:2012} and utilize a combination of a
graded-meshes, the trapezoidal quadrature rule, and a quadrature rule
that incorporates the logarithmic singularity into its quadrature
weights---as described in what follows. Interestingly, the graded
meshes and associated changes of variables gives rise to accurate
integration even in the near-singular regions mentioned above in this
section.


To introduce graded meshes we consider the polynomial change of
variables $t = w(s)$ where
\begin{eqnarray}
w(s)&=&2\pi\frac{[v(s)]^p}{[v(s)]^p+[v(2\pi-s)]^p},\quad 0\leq s\leq 2\pi,\label{eq:change_variable}\\
v(s)& =& \left(\frac{1}{p}-\frac{1}{2}\right)\left(\frac{\pi-s}{\pi}\right)^3+\frac{1}{p}\frac{s-\pi}{\pi}+\frac{1}{2},\nonumber
\end{eqnarray}
and where $p\geq 2$. The function $w$ is smooth and increasing on $[0,
2\pi]$, with $w^{(k)}(0) = w^{(k)}(2\pi)= 0$ for $1 \leq k \leq p -
1$. Using this transformation we express $K$ as
\begin{eqnarray*}
K(t,\tau)&=&K(w(s),w(\sigma)) \\
&=& K_1(w(s),w(\sigma))\log\left(4\sin^2\frac{s-\sigma}{2}\right)+\widetilde K_2(s,\sigma)
\end{eqnarray*}
where
\begin{eqnarray*}
\widetilde K_2(s,\sigma)&=&K_1(w(s),w(\sigma))\log\left(\frac{r^2(w(s),w(\sigma))}{4\sin^2\frac{s-\sigma}{2}}\right)+K_2(w(s),w(\sigma)),
\end{eqnarray*}
and where the diagonal term is given by $\widetilde
K_2(s,\sigma)=2K_1(t,t)\log(w'(s)|\nex'(t)|)+K_2(t,t)$. High-order
accurate quadrature formulae for the integral
operators~\eqref{eq:parametrized_operators} based on the
$(2n-1)$--point discretization $\sigma_j=j\pi/n$ ($1\leq j\leq 2n-1$,
corresponding to integration over the curve parametrized by
$\ney(\tau)$) at evaluation points $t=t_i=w(s_i)$ with $s_i=i\pi/q$
($1\leq i\leq 2q-1$, corresponding to evaluation of the operator
at points on the curve parametrized by $\nex(t)$) can easily be
obtained~\cite{COLTON:2012} from the expressions
\begin{equation}
  \int_0^{2\pi}f(\sigma)\de\sigma \approx\frac{\pi}{n}\sum_{j=0}^{2n-1}f(\sigma_j)\label{eq:trapezoidal_rule}
\end{equation}
and
$$
\int_0^{2\pi}f(\sigma)\log\left(4\sin^2\frac{s-\sigma}{2}\right)\de \sigma\approx \sum_{j=0}^{2n-1}R_j^{(n)}(s)f(\sigma_j),
$$
$0\leq s\leq 2\pi$, (which, for smooth functions $f$, yield
high-order accuracy), where the weights $R_j(s)$ are given by
$$
R_j(s)  = -\frac{2\pi}{ n }\sum_{m=1}^{n-1}\frac{1}{m}\cos m(s-\sigma_j)-\frac{\pi}{n^2}\cos n(s-\sigma_j).
$$
Clearly setting $s=\sigma_i$ in this equation gives $R_j(\sigma_i) =
R_{|i-j|}$ where
$$
R_k = -\frac{2\pi}{n}\sum_{m=1}^{n-1}\frac{1}{m}\cos\frac{m k \pi}{n}-\frac{(-1)^k\pi}{n^2}.
$$

Using these quadrature points and weights and corresponding parameter
values $t=t_i=w(s_i)$ for the observation point ($s_i=i\pi/q$) we
obtain the desired discrete approximation for the
integrals~(\ref{eq:parametrized_operators}): for an approximation
$\phi_j\approx \phi(\tau_j)= \phi(w(\sigma_j))$ we have
\begin{equation}
\begin{array}{c}
  \displaystyle\int_0^{2\pi}K(t_i,\tau)\phi(\tau)\de\tau  \approx\sum_{j=1}^{2n-1}\left\{K_1(t_i,\tau_j)W_{ij}+
    K_2(t_i,\tau_j)\frac{\pi}{n}\right\} \phi_j\, w'(\sigma_j)
\end{array}
 \label{eq:quadrature}
\end{equation}
for $1\leq i\leq 2q-1$, where $\tau_j = w(\sigma_j)$ and where the
quadrature weights are given by
$$
W_{ij}=R_{|i-j|}+\frac{\pi}{n}\log\left(\frac{r^2(t_i,t_j)}{4\sin^2(s_i-s_j)/2}\right).  $$
Note that for sufficiently large values of $p$ the product
$\phi(w(\sigma))w'(\sigma)$, (an approximation of which appears
in~\eqref{eq:quadrature}) vanishes continuously at the endpoints of
the parameter interval~$[0,2\pi]$---even in cases for which, as it
happens for corners or points of junction between multiple dielectric
materials, $\phi(w(\sigma))$ tends to infinity at the endpoints.

The systems of boundary integral
equations~\eqref{eq:integral_equations_I},
\eqref{eq:integral_equations_II} and \eqref{eq:integral_equations_III}
are discretized by means of applications of the quadrature
rule~\eqref{eq:quadrature} to the relevant integral
operators~\eqref{eq:integral_operators_I},
\eqref{eq:integral_operators_II} and
\eqref{eq:integral_operators_III}, respectively.  This procedure leads
to linear systems of algebraic equations for the unknown values of the
density functions $\psi_{\inte}$ and $\psi_{\exte}$ at the quadrature
points.  The presence of the weight $w'(\sigma_j)$
in~\eqref{eq:quadrature}, which multiplies the unknowns
$\phi_j\approx\phi(\tau_j)$ and which is very small for $\sigma_j$
close to $0$ and $2\pi$, however, gives rise to highly ill conditioned
linear systems.  To avoid this difficulty we resort to the change of
unknown~$\eta_j = \phi_jw'(\sigma_j)$ in \eqref{eq:quadrature}; for
the equations which contain terms of the form $\psi_\inte/2$ and
$\psi_\exte/2$ it is additionally necessary to multiply both sides of
the equation by $w'(\sigma_j)$ to avoid small denominators. In what
follows, the resulting discrete linear systems for the problems under
consideration are generically denoted by $\bold A\boldsymbol\eta =
\bold f$ where, in each case $\boldsymbol\eta$ is a vector that
combines the unknowns that result from the discretization procedure
described above in this section for the various boundary portions
$\Gamma_{ij}$ (cf. Fig.~\ref{fig:mathematical_description}). Once
$\boldsymbol\eta$ has been found, the numerical approximation of the
scattered fields at a given point $\nex$ in space, which in what
follows will be denoted by $\tilde u = \tilde u(\nex)$, can be obtained
by consideration of the relevant
representation~\eqref{eq:single_layer_representation_I},
\eqref{eq:single_layer_representation_II} or
\eqref{eq:single_layer_representation_III}. For evaluation points
$\nex$ sufficiently far from the integration curves these integrals
can be accurately approximated using the change of variable $t=w(s)$
together with the trapezoidal rule~(\ref{eq:trapezoidal_rule}); for
observation points near the integration curves, in turn, a procedure
based on interpolation along a direction transverse to the curve is
used (see~\cite{eldar:2014} for details).

\subsection{\label{sec:non_uniqueness}Solution at resonant and
  near-resonant frequencies}

As mentioned in the introduction, despite the fact that each one of
the physical problems considered in this contribution admit unique
solutions for all frequencies~$\omega$ and all physically admissible
values of the dielectric constant and magnetic permeability, for
certain values of $\omega$ spurious resonances occur: for such values
of $\omega$ the systems of integral equations derived in
Sec.~\ref{sec:integral_formulation} are not invertible. In fact,
spurious resonances for these systems arise whenever the wavenumber
$k_3$, which will also be denoted by~$\kappa$ in what follows, is such
that $-k_3^2=-\kappa^2$ equals a certain Dirichlet eigenvalue.  (More
precisely, letting $\varepsilon(x)$ and $\mu(x)$ denote the prescribed
(piece-wise constant) permittivity and permeability, spurious
resonances occur whenever $\kappa$ satisfies $\Delta u =
-\kappa^2\varepsilon(x)\mu(x) u$ in $\Omega_{1}\cup\Omega_2$ for some
nonzero function $u$ satisfying $u=0$ on $\partial
(\Omega_{1}\cup\Omega_2)$. This can be established e.g.  taking into
account ideas underlying uniqueness arguments of the type found
in~\cite[Chapter 3]{COLTON:1983}. Note, in particular, that the
values of $\kappa$ for which spurious resonances occur are necessarily
real numbers (and, thus, physically realizable), since the eigenvalues
$-\kappa^2$ are necessarily negative).

It is important to note that, in addition to the spurious resonances
mentioned above, the transmission problems considered in
Sec.~\ref{sec:scattering_problem} {\em themselves} (and, therefore the
corresponding systems of integral equations mentioned above) also
suffer from non-uniqueness for certain {\em non-physical} values of
$\kappa$ ($\Im(\kappa)< 0$) which are known as ``scattering
poles''~\cite{taylor1996}; cf. Fig.~\ref{fig:error_chebyshev} and a
related discussion below in this section.

The non-invertibility of the aforementioned continuous systems of
integral equations at a spurious-resonance or scattering-pole
wavenumber $\kappa=\kappa^*$ manifests itself at the discrete level in
non-invertibility or ill-conditioning of the system matrix $\bold
A:=\bold A(\kappa)$ for values of $\kappa$ close to
$\kappa^*$. Therefore, for $\kappa$ near $\kappa^*$ the numerical
solution of the transmission problems under consideration (which, in
what follows will be denoted by $\tilde u:=\tilde u_\kappa(\nex)$ to
make explicit the solution dependence on the parameter $\kappa$)
cannot be obtained via direct solution the linear system $\bold
A\boldsymbol\eta =\bold f$. As is known, however~\cite{taylor1996},
the solutions $u = u_\kappa$ of the continuous transmission problems
are analytic functions of $\kappa$ for all real values of
$\kappa$---including, in particular, for $\kappa$ equal to any one of
the spurious resonances mentioned above and for real values of
$\kappa$ near a scattering pole---and therefore, the approximate
values $\tilde u_\kappa(\nex)$ for $\kappa$ sufficiently far from
$\kappa^*$ can be used, via analytic continuation, to obtain
corresponding approximations around $\kappa = \kappa^*$ and even at a
spurious resonance~$\kappa = \kappa^*$. 

In order to implement this strategy for a given value of $\kappa$ it
is necessary for our algorithm to possess a capability to perform two
main tasks, namely, Task I:~Determination of whether $\kappa$ is
``sufficiently far'' from any one of the spurious resonances and
scattering poles~$\kappa^*$; and Task II:~Evaluation of analytic
continuations to a given real wavenumber $\kappa_0$ which is either
close or equal to a spurious resonance $\kappa^*$, or which lies close
to a scattering pole $\kappa^*$. Once these capabilities are available
the algorithm can be completed readily: if completion of Task I leads
to the conclusion that $\kappa$ is far from all spurious resonances
then the solution process proceeds directly via solution of the
associated system of integral equations. Otherwise, the solution
process is completed by carrying out Task~II. Descriptions of the
proposed methodologies to perform Tasks~I and~II are presented in the
following two sections.
\subsubsection{Task I:  matrix-singularity detection\label{Task_I}}
Consider a given wavenumber $\kappa'$ for which a solution to one of
the problems under consideration needs to be obtained. As discussed in
what follows, in order to determine the level of proximity of
$\kappa'$ to a spurious resonance or scattering pole $\kappa^*$, the
matrix-singularity detection algorithm utilizes the minimum singular
value $\sigma_{\min}(\kappa')$ of $\bold A(\kappa')$. (Note that in
view of the discussion concerning Task~I above in the present
Sec.~\ref{sec:non_uniqueness} it is not necessary to differentiate
wavenumbers $\kappa'$ that lie near to either a spurious resonance or
to a scattering pole: both cases can be treated equally well by means
of one and the same Task~II (analytic continuation) algorithm
(Sec.~\ref{anal_cont}).

To introduce the matrix-singularity detection algorithm consider
Fig.~\ref{fig:sigma_min}: clearly, with exception of a sequence of
wavenumbers (spurious resonances and/or real wavenumbers close to
non-real scattering pole) around which the minimum singular value is
small, the function $\sigma_{\min}(\kappa)$ maintains an essentially
constant level. This property forms the basis of the
matrix-singularity detection algorithm. Indeed, noting that there are
no singularities for~$\kappa$ smaller than certain threshold (as it
follows from the spectral theory for the Laplace operator), we choose
a wavenumber $\kappa_0>0$ close to zero and we compare
$\sigma_{\min}(\kappa_0)$ with $\sigma_{\min}(\kappa')$. If
$\sigma_{\min}(\kappa') \ll \sigma_{\min}(\kappa_0)$, say
$\sigma_{\min}(\kappa')< \xi \cdot\sigma_{\min}(\kappa_0)$ for an
adequately chosen value of $\xi$, $\kappa'$ is determined to be close
to a some singularity $\kappa^*$, and therefore the Task-II
analytic-continuation algorithm is utilized to evaluate $\tilde
u_{\kappa'}(\nex)$.  The parameter values $\kappa_0=0.1$ and $\xi =
10^{-4}$ were used in all the numerical examples presented in this
paper.

(A remark is in order concerning the manifestations of resonances and
scattering poles on the plots of the function $\sigma_{\min}(\kappa)$
as a function of the real variable $\kappa$. By definition the
function $\sigma_{\min}(\kappa)$ vanishes exactly at all spurious
resonances. The four sharp peaks shown in Fig.~\ref{fig:sigma_min},
for example, occur at the spurious resonances listed in the inset of
Fig.~\ref{fig:error_chebyshev}. The first peak from the left in
Fig.~\ref{fig:sigma_min}, in contrast, is not sharp---as can be seen
in the inset close-up included in the figure.  The small value
$\sigma_{\min}(\kappa)\sim 10^{-7}$ around $\kappa=0.5708$ is
explained by the presence of a scattering pole $\kappa^*$:
$\sigma_{\min}(\kappa^*)=0$ at the complex wavenumber $\kappa^* =
0.57807113743881-0.000074213015953i$. Thus scattering poles can in
practice be quite close to the real $\kappa$ axis, and thus give rise
to rather sharp peaks which are not associated with actual spurious
resonances. As mentioned above, however, the analytic continuation
algorithm presented in what follows need not differentiate between
these two types of singularities: analytic continuation is utilized
whenever a sufficiently small value of $\sigma_{\min}$ is detected.)

\subsubsection{Task II:  analytic continuation\label{anal_cont}}
Analytic continuation of the numerical solution $\tilde
u_{\kappa}(\nex)$ to a given wavenumber $\kappa'$ detected as a matrix
singularity (Sec.~\ref{Task_I}) is carried out via
interpolation. Note, however, that, since $\mathbf{A}(\kappa)$ is
generally extremely ill-conditioned for values of $\kappa$ in a narrow
interval around such wavenumbers~$\kappa'$, fine interpolation meshes
cannot be utilized to achieve arbitrary accuracy in the
approximation. To overcome this difficulty we utilize an interpolation
method based on use of Chebyshev expansions, for which the meshsize is
not allowed to be smaller than a certain tolerance, and within which
convergence is achieved, in view of the analyticity of the scattered
field with respect to the wavenumber $\kappa$, by increasing the order
of the Chebyshev expansion. To do this for a given wavenumber
$\kappa'$ identified by the matrix-singularity detection algorithm
(Sec.~\ref{Task_I}), the analytic continuation algorithm proceeds
by introducing a Chebyshev grid of points $\{\kappa_j\}_{j=1}^{2m}$
(cf.~\cite{isaacson:1994}) sorted in ascending order such that the two
middle points in the grid, $\kappa_m$ and $\kappa_{m+1}$, lie at an
appropriately selected distance $\delta>0$ from the wavenumber
$\kappa'$: $\kappa_m=\kappa'-\delta$ and
$\kappa_{m+1}=\kappa'+\delta$.

The accuracy of the numerical evaluation of the field $\tilde
u_{\kappa_j}$ at each one of the interpolation points $\kappa_j$ is
ensured by running the matrix-singularity detection algorithm at each
$\kappa_j$ and adequately changing the value of $\delta$ if a
matrix-singularity is detected at one or more of the mesh points
$\kappa_j$.  Letting $\tilde u^{(m)}_\kappa$ denote the Chebyshev
expansion of order $2m-1$ resulting for a Chebyshev mesh selected as
indicated above, the sequence $\tilde u^{(m)}_{\kappa'}$ convergences
exponentially fast to $\tilde u_{\kappa'}$ as~$m$ grows---as it befits
Chebyshev expansions of analytic functions. If the matrix-singularity
condition
$\sigma_{\min}(\kappa_{j_{\ell}})<\xi\cdot\sigma_{\min}(\kappa_0)$
occurs at one of more of the interpolation points $\kappa_{j}$, say
$\kappa_{j_\ell}$, $1\leq \ell\leq L'$, the algorithm proceeds by
selecting the smallest value of the parameter $\delta'>\delta$ and a
new set of Chebyshev points $\{\kappa'_j\}_{j=1}^{2 m'}$ ($ m'\geq m
$) satisfying $ \kappa'_{m'}=\kappa'-\delta'$, $
\kappa'_{m'+1}=\kappa'+\delta'$, such that none of the new
interpolation points lie on the region
$\bigcup_{\ell=1}^{L'}(\kappa_{j_\ell}-\delta,\kappa_{j_{\ell}}+\delta)$.
If the condition
$\sigma_{\min}(\kappa'_{j})<\xi\cdot\sigma_{\min}(\kappa_0)$ occurs
for some of the new interpolation points, say $\kappa'_{j_\ell}$,
$1\leq \ell\leq L''$, the algorithm proceeds as described above, but
for a new value $ \delta''> \delta'$, and so on. Note that in practice
the interpolation procedure described above is rarely needed, and when
it is needed, a suitable interpolation grid is usually found after a
single iteration: in practice the choice $\delta = 0.01$ has given
excellent results in all the examples presented in this paper.

\begin{figure}[h!]	
\centering	
\includegraphics[width=10cm]{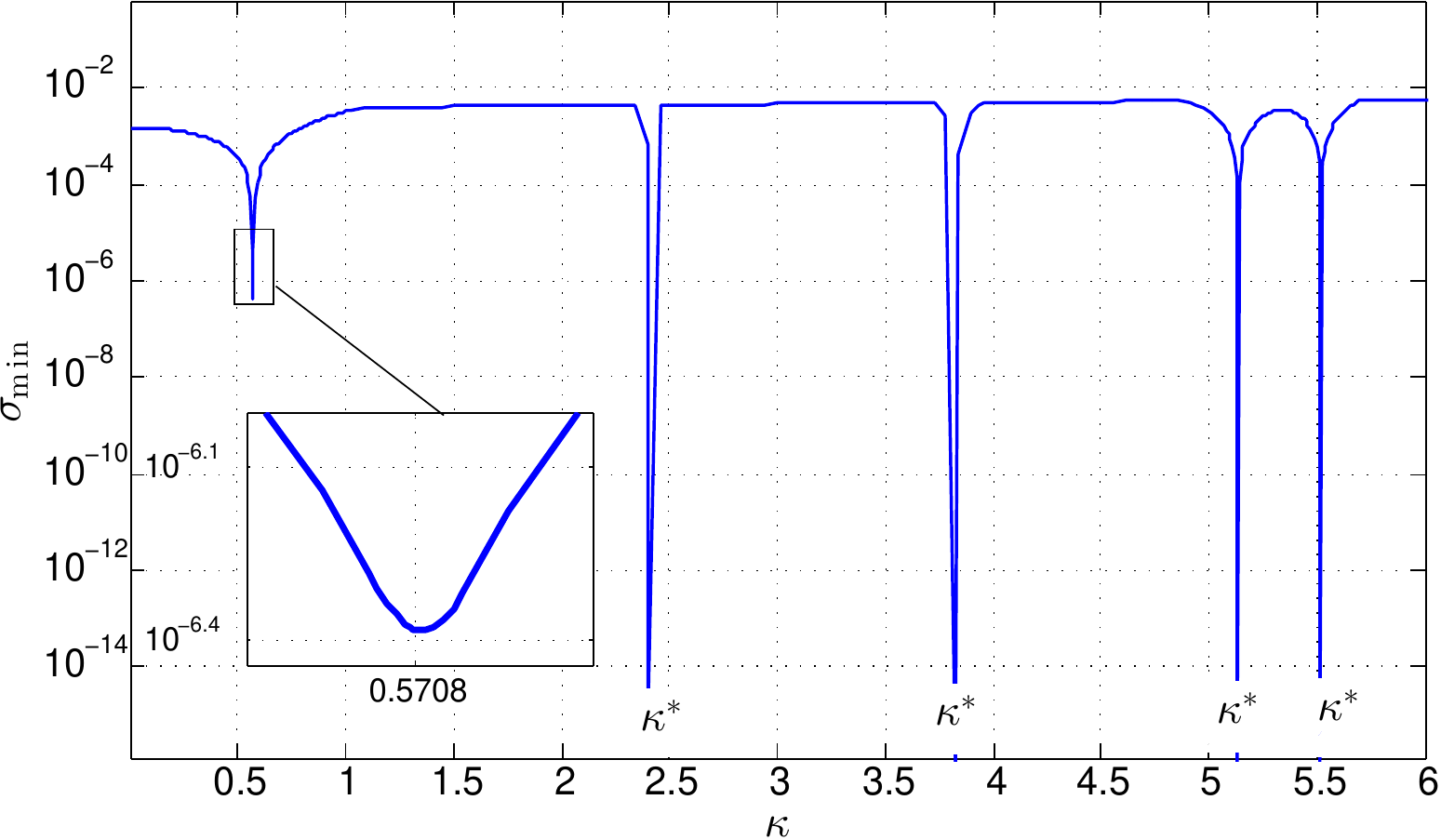}
\caption{Minimum singular value of $\bold A$ as a function of~$\kappa = k_3$ for the problem of scattering by a semi-circular bump on a PEC half-plane in TE-polarization.}\label{fig:sigma_min}
\end{figure}

\begin{figure}[h!]	
\centering	
\includegraphics[width=10.2cm]{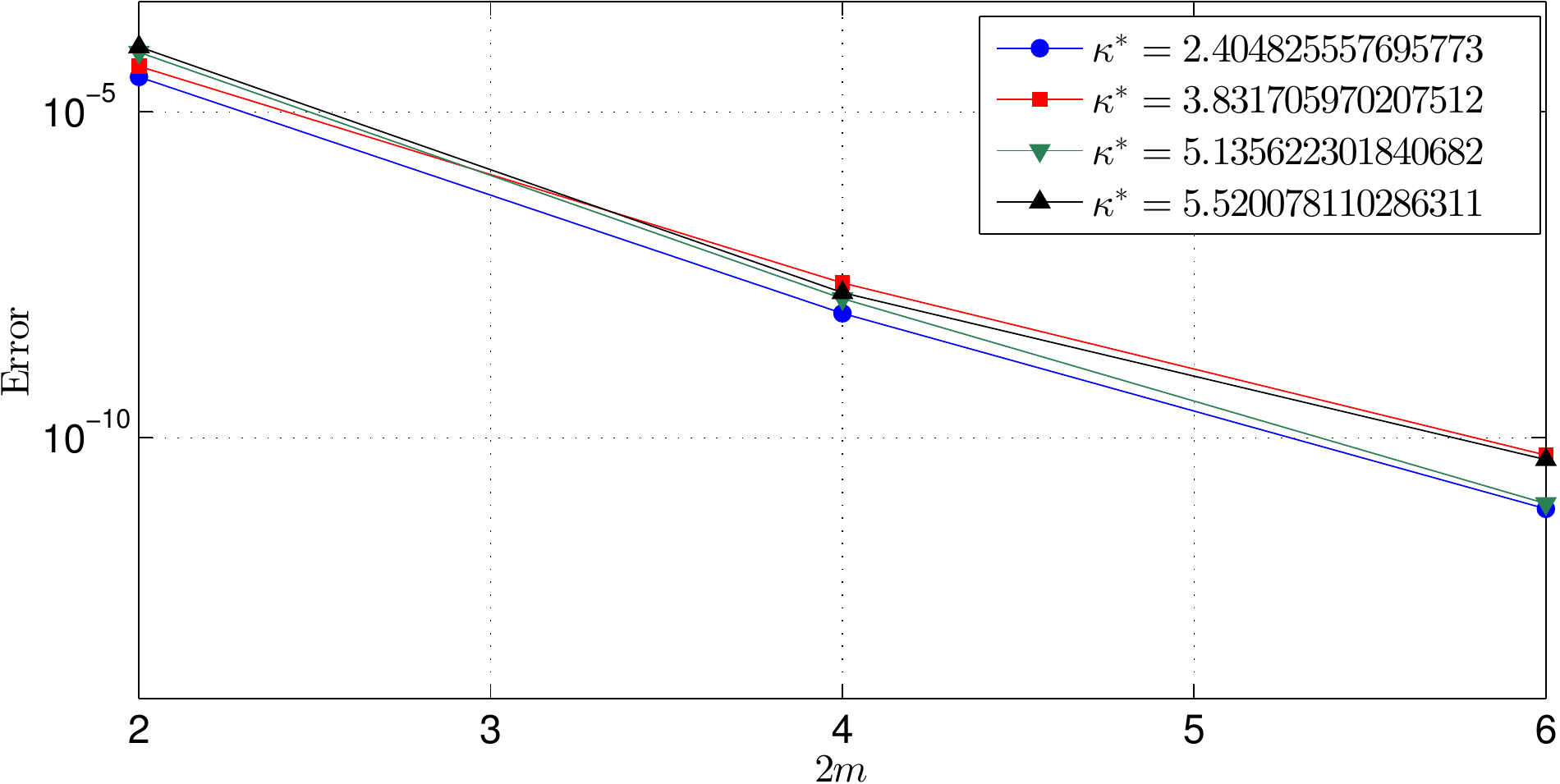}
\caption{Error in the approximation of $\tilde u_{\kappa^*}$ by
  Chebyshev interpolation/analytic-continuation for various spurious
  resonant frequencies $\kappa^*$ as a function of the order $2m$ of
  the Chebyshev expansion.}\label{fig:error_chebyshev}
\end{figure}
In order to demonstrate the fast convergence of $\tilde
u^{(m)}_{\kappa'}$ to $\tilde u_{\kappa'}$ as $m$ increases we
consider the problem of scattering by a dielectric unit-radius
semi-circular bump on a PEC half-plane. For this problem the
wavenumbers~$\kappa^*$ for which the system of integral
equations~\eqref{eq:integral_equations_III} is non-invertible can be
computed explicitly: spurious resonances are given by real solutions
of the equation~{\small $J_n(\kappa)=0$}, $n\geq0$, where~{\small
  $J_n$} denotes the Bessel function of first kind and order~$n$, and
scattering poles are complex valued solutions of {\small $\kappa
  H_n^{(1)}(\kappa)J_n'(k_1)=k_1J_n(k_1){H_n^{(1)}}'(\kappa)$}, where
{\small $H_n^{(1)}$} denotes the Hankel function of first kind and
order $n$
(see~Appendix.~\ref{app:bump}\ref{sec:scattering_poles}). The function
$\sigma_{\min}(\kappa)$ is displayed in Fig.~\ref{fig:sigma_min}. The
$\kappa^*$ values identified in that figure coincide (up to machine
precision) with the first four positive solutions of the
equation~{\small $J_n(\kappa)=0$}. On the other hand, this problem
admits an analytical solution $u_{\kappa}$ in terms of a
Fourier-Bessel expansion~(see Appendix~\ref{app:bump}). The
availability of the exact solution allows us to quantify the magnitude
of interpolation error by evaluating the maximum of the error function
$E(\nex) = |\tilde u^{(m)}_{\kappa^*}(\nex)- u_{\kappa^*}(\nex)|$ at a
polar grid $\Pi$ (consisting of points inside, outside and at the
boundary of the semi-circular bump). Fig.~\ref{fig:error_chebyshev}
shows the error $\max_{\nex\in \Pi} E(\nex)$ versus the number of
points used in the Chebyshev interpolation of $\tilde u_{\kappa^*}$,
which is computed for the four spurious resonances $k^*$ shown in
Fig.~\ref{fig:sigma_min}, and where a sufficiently fine spatial
discretization is used. In all the calculations $k_1 = 6$, the curve
$\Gamma_{13}$ is discretized using $128$ points, and $\delta = 0.01$
is utilized to construct the Chebyshev grids.


\section{\label{sec:numerical_results}Numerical results}
This section demonstrates the high accuracies and high-order
convergence that result as the proposed boundary integral methods are
applied to each one of the mathematical problems formulated in
Sec.~\ref{sec:integral_formulation}.  For definiteness all dielectric
media are assumed non-magnetic so that $\beta_i/\beta_j=1$ for
TM-polarization and $\beta_i/\beta_j=k_i^2/k_j^2$ for
TE-polarization. In all the numerical examples shown in this section
the incident plane-wave is parallel to the vector
$d=(\cos(\pi/3),-\sin(\pi/3))$ and the graded-mesh
parameter~\eqref{eq:change_variable} is $p=8$.
\begin{figure}[h!]

	\centering
	\begin{subfigure}[b]{0.3\textwidth}		   		
   	\centering	
	\includegraphics[width=\textwidth]{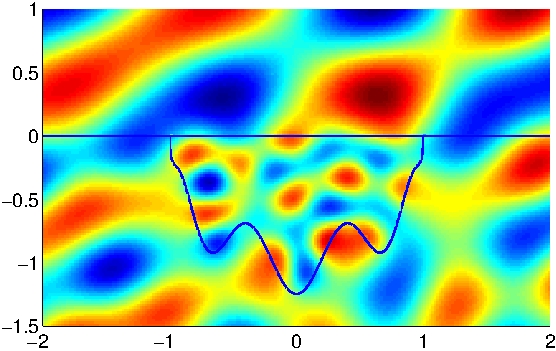}
    \caption{TM-polarization.}\label{I:TM-polarization}
	\end{subfigure} \qquad 
    \centering       
    \hspace{0.0cm}\begin{subfigure}[b]{0.3\textwidth}     		
    \includegraphics[width=\textwidth]{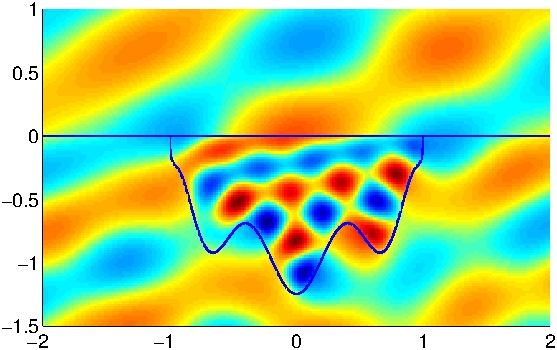}
    \caption{TE-polarization.}\label{I:TE-polarization}
	\end{subfigure}	\bigskip\\
	
	\centering
	\begin{subfigure}[b]{0.3\textwidth}		   		
   	\centering	
	\includegraphics[width=\textwidth]{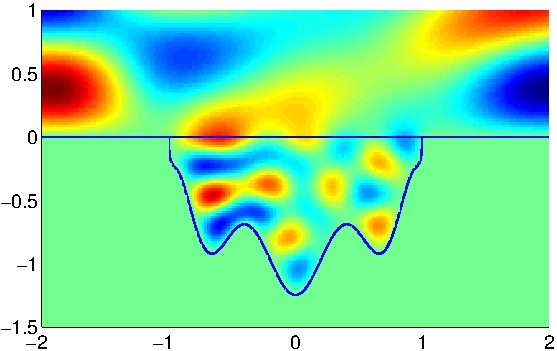}
	\caption{TM-polarization.}\label{II:TM-polarization}
	\end{subfigure} \qquad   	
    	\centering       
    \hspace{0.0cm}\begin{subfigure}[b]{0.3\textwidth}     		
    \includegraphics[width=\textwidth]{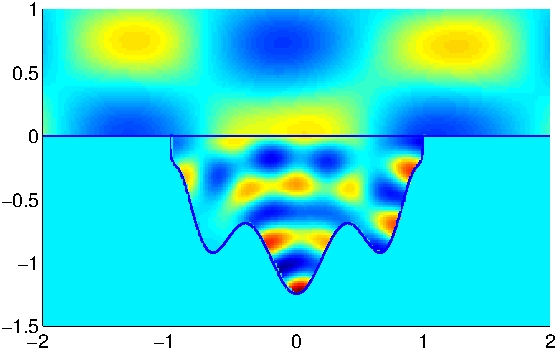}
	\caption{TE-polarization.} \label{II:TE-polarization}
	\end{subfigure}	\bigskip\\

	\centering
	\begin{subfigure}[b]{0.3\textwidth}		   		
   	\centering	
	\includegraphics[width=\textwidth]{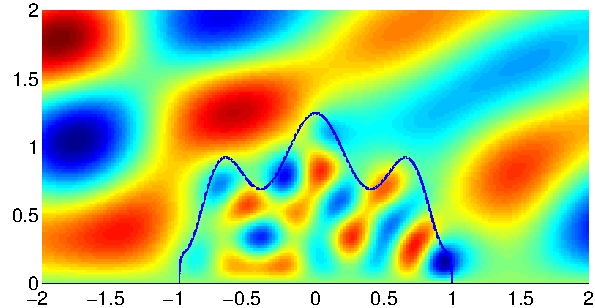}
	\caption{TM-polarization.}\label{III:TM-polarization}
	\end{subfigure} \qquad
    \centering       
    \hspace{0.0cm}\begin{subfigure}[b]{0.3\textwidth}     		
    \includegraphics[width=\textwidth]{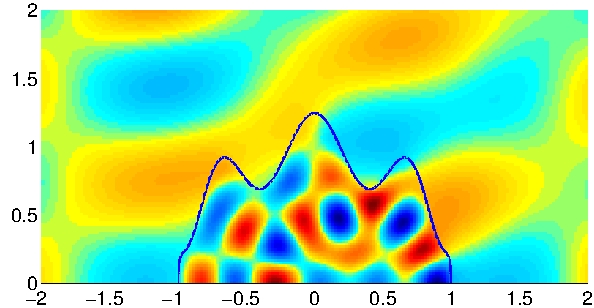}
	\caption{TE-polarization.}\label{III:TE-polarization}
	\end{subfigure}	
	
    \caption{Diffraction pattern resulting from the scattering of a plane-wave by; a dielectric-filled cavity on a dielectric half-plane ((a) and (b)); a~dielectric-filled cavity on a PEC half-plane ((c) and (d)); a dielectric bump on a PEC half-plane ((e)~and~(f)).}\label{fig:diffraction_patern}
\end{figure}

\begin{figure}[h!]
\centering
	\begin{subfigure}[b]{0.3\textwidth}		   		
   	 	\centering	
\includegraphics[width=\textwidth]{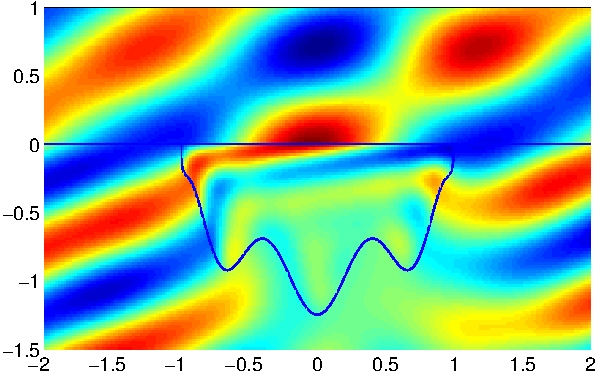}
		\caption{TM-polarization.}\label{I:TM-polarization_complex}
	\end{subfigure} \qquad
        \centering       
    \hspace{0.0cm}\begin{subfigure}[b]{0.3\textwidth}     		
        \includegraphics[width=\textwidth]{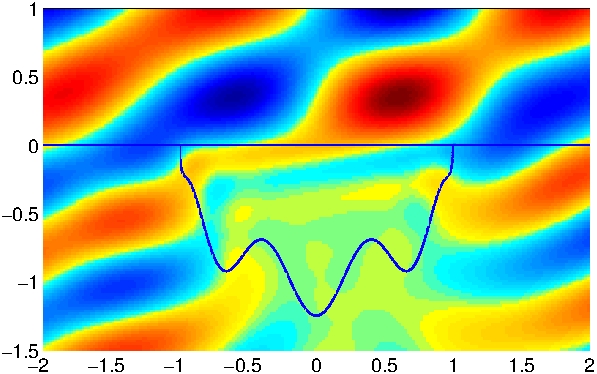}
        \caption{TE-polarization.}\label{I:TE-polarization_complex}
	\end{subfigure}	 \bigskip\\
\centering
	\begin{subfigure}[b]{0.3\textwidth}		   		
   	 	\centering	
\includegraphics[width=\textwidth]{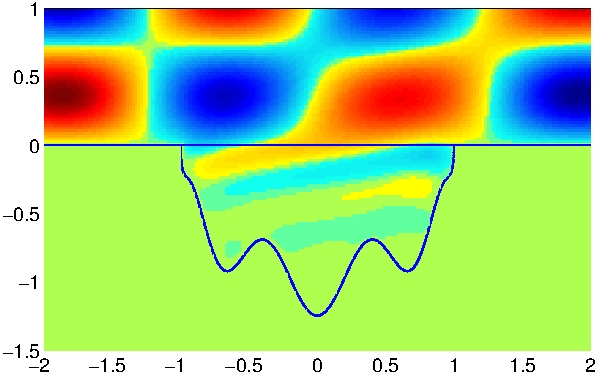}
		\caption{TM-polarization.}\label{II:TM-polarization_complex}
	\end{subfigure} \qquad
       \centering       
    \hspace{0.0cm}\begin{subfigure}[b]{0.3\textwidth}     		
        \includegraphics[width=\textwidth]{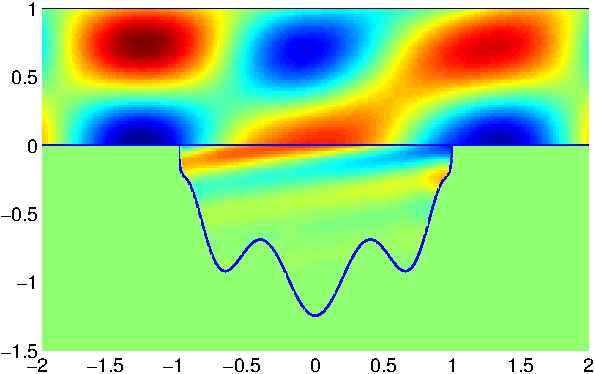}
        \caption{TE-polarization.} \label{II:TE-polarization_complex}
	\end{subfigure}	\bigskip\\
\centering
	\begin{subfigure}[b]{0.3\textwidth}		   		
   	 	\centering	
\includegraphics[width=\textwidth]{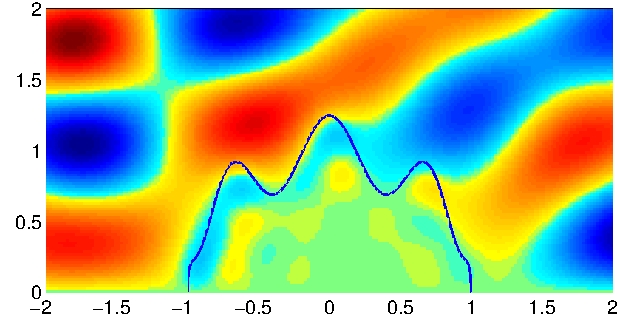}
		\caption{TM-polarization.}\label{III:TM-polarization_complex}
	\end{subfigure} \qquad
        \centering       
    \hspace{0.0cm}\begin{subfigure}[b]{0.3\textwidth}     		
        \includegraphics[width=\textwidth]{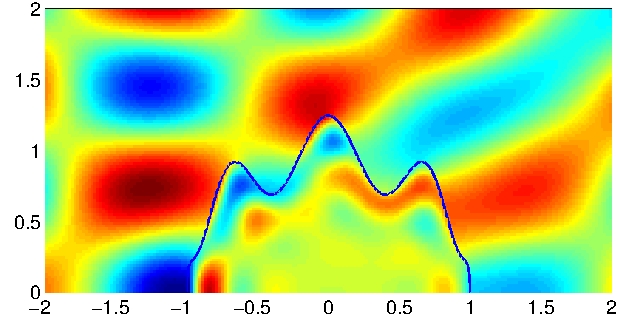}
        \caption{TE-polarization.}\label{III:TE-polarization_complex}
	\end{subfigure}	
    \caption{Diffraction pattern resulting from the scattering of a plane-wave by; a dielectric-filled cavity on a dielectric half-plane ((a) and (b)); a~dielectric-filled cavity on a PEC half-plane ((c) and (d)); a dielectric bump on a PEC half-plane ((e)~and~(f)).}\label{fig:diffraction_patern_complex}
\end{figure} 
We thus consider the problem of scattering by a dielectric filled
cavity on a dielectric half-plane (problem type I); the problem of
scattering by a dielectric filled cavity on a PEC half-plane (problem
type II); and the problem of scattering by a dielectric bump on a PEC
half-plane (problem type III). With reference to
Fig.~\ref{fig:mathematical_description}, in the first two examples
the cavity is determined by the curve $\Gamma_{24} = \{(x,y)\in\R^2:x
= -\cos(\frac{t}{2}),y=\frac{\cos(4t)}{40}t(t-2\pi)-\sin(\frac{t}{2}),
t\in(0,2\pi)\}$, and the curve $\Gamma_{13}$ (which, in view of the
formulation in Sec.~\ref{sec:integral_formulation}, may be selected
rather arbitrarily as long as it lies in the upper half plane and has
the same endpoints as $\Gamma_{24}$) is given by the semicircle
of radius one in the upper half plane that joins the points $(1, 0)$
and $(-1, 0)$. For the problem of scattering by a dielectric bump (type
III problem), in turn, the boundary of the bump is given by
$ \Gamma_{13}= \{(x,y)\in\R^2:x =
\cos(\frac{t}{2}),y=\frac{\cos(4t)}{40}t(2\pi-t)+\sin(\frac{t}{2}),
t\in(0,2\pi)\}.  $

To estimate the error in the aforementioned numerical test problems,
the systems of boundary integral equations
\eqref{eq:integral_equations_I}, \eqref{eq:integral_equations_II} and
\eqref{eq:integral_equations_III} were discretized utilizing five
different meshes $\Pi_j$, $1\leq j\leq 5$ consisting of $P=2^{j+5}-1$
points distributed along each one of the relevant boundaries: $P$
points on $\Gamma_{24}$ and $P$ points on $\Gamma_{13}$ in the case of
type I and II problems, and $P$ points on $\Gamma_{13}$ in the case of
type III problem. The sequence of meshes is chosen to be nested
($\Pi_j\subset\Pi_i$ for $j<i$) in order to facilitate the convergence
analysis; in what follows the numerical solution that results from the
discretization $\Pi_j$ is denoted by $\tilde u_j$.  The error in the
numerical solution $\tilde u_j$ is estimated by means of the
expression
$$
E_j = \frac{\max_{\nex\in \Pi_1}|\tilde u_j(\nex)-\tilde
  u_5(\nex)|}{\max_{\nex\in \Pi_1}|\tilde u_5(\nex)|},\quad 1\leq
j\leq 4.
$$

\begin{table}[h!]
\begin{center}
    \begin{tabular}{c|c|c|c|c|c|c|c}
         \multicolumn{2}{c|}{  } &\multicolumn{2}{c|}{ Type I } & \multicolumn{2}{c|}{Type II} &\multicolumn{2}{c}{ \rule{0pt}{2.3ex}    Type III}  \\\hline
        \multicolumn{2}{c|}{  } &\multicolumn{2}{c|}{ $k_2$}&\multicolumn{2}{c|}{ $k_2$} &\multicolumn{2}{c}{$k_1$}  \\ \hline

 \rule{0pt}{2.3ex}           & $P$  & $15$     & $15+5i$  & $15$ & $15+5i$  & $15$ &$15+5i$   \\ 
    \hline 
 \rule{0pt}{2.5ex}        
&    63 &  $3\!\cdot\! 10^{-01}$  & $6\!\cdot\! 10^{-03}$ & 
       	   $7\!\cdot\! 10^{-01}$  & $1\!\cdot\! 10^{-04}$ &
       	   $2\!\cdot\! 10^{-01}$  & $7\!\cdot\! 10^{-02}$ \\
\footnotesize{TM}   
&   127 &  $7\!\cdot\! 10^{-04}$ & $4\!\cdot\! 10^{-06}$ &
           $2\!\cdot\! 10^{-03}$ & $1\!\cdot\! 10^{-07}$ &
           $2\!\cdot\! 10^{-03}$ & $1\!\cdot\! 10^{-03}$ \\
&   255 &  $1\!\cdot\! 10^{-10}$ & $7\!\cdot\! 10^{-12}$ & 
		   $3\!\cdot\! 10^{-11}$ & $6\!\cdot\! 10^{-12}$ &
		   $5\!\cdot\! 10^{-08}$ & $8\!\cdot\! 10^{-08}$ \\
&   511 &  $6\!\cdot\! 10^{-12}$ & $5\!\cdot\! 10^{-12}$ & 
           $1\!\cdot\! 10^{-12}$ & $3\!\cdot\! 10^{-13}$ &
           $1\!\cdot\! 10^{-13}$ & $8\!\cdot\! 10^{-13}$ \\
 \hline
 \rule{0pt}{2.5ex}   
& 63    &  $9\!\cdot\! 10^{-02}$ & $3\!\cdot\! 10^{-03}$ & 
           $2\!\cdot\! 10^{-01}$ & $6\!\cdot\! 10^{-04}$ &
           $4\!\cdot\! 10^{-01}$ & $4\!\cdot\! 10^{-02}$ \\
\footnotesize{TE} 
&  127  &  $3\!\cdot\! 10^{-04}$ & $7\!\cdot\! 10^{-06}$ & 
	       $1\!\cdot\! 10^{-04}$ & $2\!\cdot\! 10^{-07}$ &
	       $1\!\cdot\! 10^{-03}$ & $3\!\cdot\! 10^{-04}$ \\
&  255  &  $3\!\cdot\! 10^{-12}$ & $2\!\cdot\! 10^{-12}$ & 
           $3\!\cdot\! 10^{-12}$ & $7\!\cdot\! 10^{-12}$ & 
           $2\!\cdot\! 10^{-08}$ & $2\!\cdot\! 10^{-08}$ \\
&   511 &  $1\!\cdot\! 10^{-12}$ & $2\!\cdot\! 10^{-12}$ & 
           $4\!\cdot\! 10^{-14}$ & $1\!\cdot\! 10^{-14}$ &
           $1\!\cdot\! 10^{-13}$&  $2\!\cdot\! 10^{-13}$\\ 
\hline
    \end{tabular} 
  \caption{Convergence test for the numerical solution of Problem Type~I ($k_1=5$, $k_2=15$ or $15+5i$, $k_3=5$, and $k_4=7$), II ($k_1=5$, $k_2=15$ or $15+5i$, and $k_3=5$) and III ($k_1=15$ or $15+5i$, and  $k_3=5$).\label{tb:table_1}}
  \end{center}
\end{table}

Table~\ref{tb:table_1} presents the numerical error estimates $E_j$,
$1\leq j\leq 5$ for the three different problem types (including real
and complex wavenumbers); clearly high accuracies and fast convergence
is achieved in all cases.  To further illustrate the results provided
by the proposed method, the real part of the total field is presented
in Figs.~\ref{fig:diffraction_patern}
and~\ref{fig:diffraction_patern_complex} for the cases considered in
Table~\ref{tb:table_1}, including examples for TM- and
TE-polarization.  Thus,
Figs.~\ref{I:TM-polarization}-\ref{I:TE-polarization} ($k_2=15$) and
Figs.~\ref{I:TM-polarization_complex}-\ref{I:TE-polarization_complex}
($k_2=15+5i$) present the diffraction pattern for the problem of
scattering by the dielectric-filled cavity on the dielectric
half-plane (problem Type I);
Figs.~\ref{II:TM-polarization}-\ref{II:TE-polarization} ($k_2=15$) and
Figs.~\ref{II:TM-polarization_complex}-\ref{II:TE-polarization_complex}
($k_2=15+5i$) present the diffraction pattern for the problem of
scattering by the dielectric-filled cavity on the PEC half-plane
(problem Type II); and
Figs.~\ref{III:TM-polarization}-\ref{III:TE-polarization} ($k_1=15$)
and
Figs.~\ref{III:TM-polarization_complex}-\ref{III:TE-polarization_complex}
($k_1=15+5i$) present the diffraction pattern for the problem of
scattering by the dielectric bump on the PEC half-plane.

\begin{figure}
\begin{center}
	\begin{subfigure}[b]{0.3\textwidth}		   		
\includegraphics[width=\textwidth]{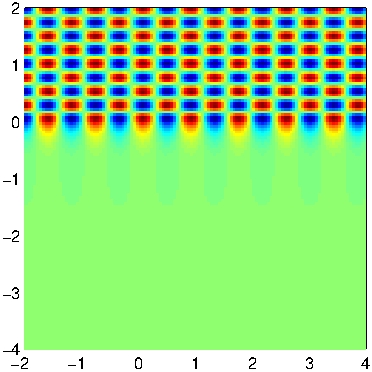}
		\caption{TM-polarization.}\label{III:TM-polarization_pw}

	\end{subfigure}\qquad
    \hspace{0.0cm}\begin{subfigure}[b]{0.3\textwidth}     		
        \includegraphics[width=\textwidth]{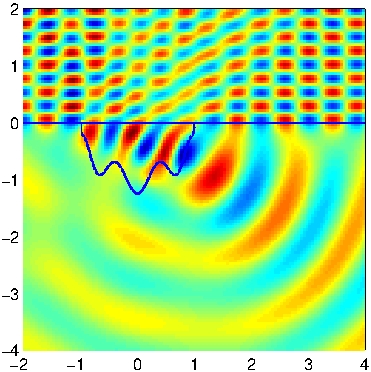}
        \caption{TM-polarization.}\label{III:TE-polarization_pw}
	\end{subfigure}		\bigskip\\
\centering
	\begin{subfigure}[b]{0.3\textwidth}		   		
\includegraphics[width=\textwidth]{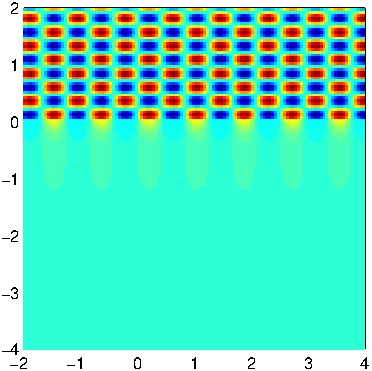}
		\caption{TE-polarization.}\label{III:TM-polarization_int}

	\end{subfigure}\qquad
    \hspace{0.0cm}\begin{subfigure}[b]{0.3\textwidth}     		
        \includegraphics[width=\textwidth]{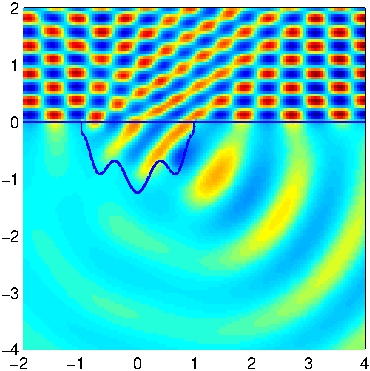}
        \caption{TE-polarization.}\label{III:TE-polarization_int}
	\end{subfigure}	
\end{center}
\caption{Scattering and transmission of a plane-wave by a dielectric
  half plane in absence (resp. presence) of a dielectric-filled cavity
  (Problem Type I with wavenumbers $k_1=k_3=15$, $k_2=10$ and
  $k_4=5$. The parameters are selected in such a way to give rise to
  total internal reflection in absence of the cavity.}
\label{fig:diffraction_patern_interesting}
\end{figure}

Fig.~\ref{fig:diffraction_patern_interesting}, finally, presents
diffraction patterns (real part) for the problem of scattering by a
dielectric filled cavity on a dielectric half-plane (Problem Type I)
for the wavenumbers $k_1=k_3=15$, $k_2=10$, $k_4=5$ and the angle of
incidence $\alpha = -\pi/3$ in TM- and TE-polarization, as well as the
corresponding transmission patterns for the dielectric half-plane in
absence of the cavity. For these specially selected numerical values
of the physical constants the phenomenon of total internal
reflection~\cite{JACKSON:1998} takes place: in absence of the cavity
the field transmitted below the interface decays exponentially fast
with the distance to the interface. Interestingly (although not
surprisingly), placement of a defect in this configuration gives rise
to transmission of electromagnetic radiation to the lower half plane.  
\newline

\noindent {\bf Acknowledgments.} The authors gratefully acknowledge
support from the Air Force Office of Scientific Research and the
National Science Foundation.

\appendix
\section{Semi-circular dielectric bump on PEC half-plane\label{app:bump}\label{sec:exact_solution}}
For reference and testing we consider the problem of scattering of a
plane-wave by a unit-radius semi-circular dielectric bump $\Omega_1 =
\{(r\cos\theta,r\sin\theta)\in \R^2,0\leq r<1, 0\leq \theta\leq \pi\}$
on a PEC half-plane $\Omega_4$ (Problem Type III), for which an exact
solution in terms of a Fourier-Bessel expansion exists. In detail, the
solution of \eqref{eq:helmholtz_equation} can expressed as
\begin{subequations}\begin{eqnarray}
    u_1(r,\theta) &=& \sum_{n=0}^{\infty}a_n J_{n}(k_1r) \Phi_n(\theta)\quad\mbox{in}\quad\Omega_1,\\
    u_3(r,\theta) &=& \sum_{n=0}^{\infty}b_n H_{n}^{(1)}(k_3r)\Phi_n(\theta)\quad\mbox{in}\quad\Omega_3,\end{eqnarray}\label{eq:bump_exact_solution}\end{subequations} where $J_n$ and $H^{(1)}_n$ are the Bessel and Hankel functions of the first kind and order 
$n$, where $\Phi_n(\theta) = \cos(n\theta)$ in TE-polarization and  
$\Phi_n(\theta) = \sin(n\theta)$ in  TM-polarization. The  Fourier coefficients in~\eqref{eq:bump_exact_solution} are given by 
\begin{eqnarray*}
a_n&=& \frac{ c_n k_3+(d_n-nc_n ) A_n}{\left[\frac{\beta_3}{\beta_1}k_1J'_{n}(k_1)-n J_n( k_1)\right] A_n+k_3J_{n}(k_1)},\\
b_n&=&-\frac{c_n k_1 +(\frac{\beta_1}{\beta_3} d_n-nc_n) B_n }{\left[\frac{\beta_1}{\beta_3} k_3 {H^{(1)}_{n}}'(k_3)-n H^{(1)}_n(k_3)\right]B_n+k_1H^{(1)}_{n}(k_3) },
\end{eqnarray*}
where $$A_n = \frac{H^{(1)}_n(k_3)}{H^{(1)}_{n+1}(k_3)},\quad B_n = \frac{J_{n}(k_1)}{J_{n+1}(k_1)},$$ 
\begin{eqnarray*}
c_n&=&\frac{2}{(1+\delta_{n0})\pi}\int_{0}^{\pi}f(1,\theta)\Phi_n(\theta)\de\theta,\\
d_n&=&\frac{2}{(1+\delta_{n0})\pi}\int_{0}^{\pi}\frac{\p f}{\p r}(1,\theta)\Phi_n(\theta)\de\theta.
\end{eqnarray*}
\section{Scattering poles{\label{sec:scattering_poles}}}
As discussed in Sec.~\ref{sec:non_uniqueness}, scattering poles are
complex wavenumbers $k$ for which there exists a non-trivial solution
of a transmission problem without sources. In the context of the
problem of a dielectric bump on a PEC half plane, for example,
scattering poles correspond to existence of non-zero solutions of
Problem Type III with $f=0$. In the particular case considered in
Appendix~\ref{sec:exact_solution} (semi-circular bump), the problem of
evaluation of scattering poles can be further reduced to the problem
of finding zeroes of certain nonlinear equations. Indeed, in order
for~$k_3$ to be a scattering pole the conditions $$u_1 = u_3\quad
\mbox{and}\quad \frac{1}{\beta_1}\frac{\p u_1}{\p
  r}=\frac{1}{\beta_3}\frac{\p u_3}{\p r}$$ must be satisfied on the
boundary $r=1$ of the bump. From eq.~\eqref{eq:bump_exact_solution} it
follows that $k_3$ is a scattering pole if and only if there exist
non-trivial constants $a_n$ and $b_n$ such that
\begin{eqnarray*}
a_nJ_n(k_1)-b_nH_n^{(1)}(k_3) &=& 0,\\
a_n\frac{k_1J'_n(k_1)}{\beta_1}-b_n\frac{k_3{H^{(1)}_n}'(k_3)}{\beta_3}&=&0
\end{eqnarray*}
for some non-negative integer $n$. Clearly such
constants exist if and only if the determinant of the matrix
associated to the linear system above vanishes at $k_3$. Therefore,
scattering poles are given by complex valued solutions $k_3$ of the
equation
\begin{eqnarray*}
\frac{k_1}{\beta_1}H_n^{(1)}(k_3)J_n'(k_1)&=&\frac{k_3}{\beta_3}J_n(k_1){H_n^{(1)}}'(k_3)
\end{eqnarray*}  
for some non-negative integer $n$.

\section{Green function for a two-layer medium: Sommerfeld integrals\label{app:greens_function}}
Consider the Helmholtz equation in the regions
$\R^2_+=\{(x_1,x_2)\in\R^2,x_2>0\}$ and
$\R^2_-=\{(x_1,x_2)\in\R^2,x_2<0\}$ with respective wavenumbers 
$k_+$ and $k_-$. The Green function of the problem
satisfies:
\begin{equation}
\begin{array}{rccllll}
\Delta_{\nex} G+k_\pm^2 G&=&-\delta_{\ney}&\mbox{in}&\R^2_\pm,\smallskip\\
\left[G\right] &=&0&\mbox{on}& \{x_2=0\}, \smallskip\\
\displaystyle\left[\frac{1}{\beta}\frac{\p G}{\p x_2}\right] &=&0 &\mbox{on}&\{x_2=0\},
\end{array}\label{eq:def_greens_function}
\end{equation}
and the Sommerfeld radiation condition at infinity, where $\delta_{\ney}$ denotes the Dirac delta function centered at the
point $\ney\in\R^2$. As is known $G$ can be computed explicitly in
terms of Sommerfeld integrals. To obtain such explicit expressions,
given a fixed point $\ney$ we define the functions
$\varphi_\pm(\nex)=G(\nex,\ney)$, $\nex\in\R^2_\pm$. Expressing
$\varphi_\pm$ as inverse Fourier transforms
\begin{equation}
\varphi_\pm(x_1,x_2) = \frac{1}{\sqrt{2\pi}}\int_{-\infty}^{\infty} \hat\varphi_\pm(\xi,x_2)\e^{i\xi x_1}\de x_1\label{eq:fourier_transform}
\end{equation} and replacing \eqref{eq:fourier_transform} in \eqref{eq:def_greens_function} a system of ordinary differential equations for the unknown functions $\hat\varphi_\pm$ is obtained which can be solved analytically. 
Two cases arise. For $\ney\in \R^2_+$, the solution of the ODE system is given by 
\begin{equation*}
\begin{array}{rcl}
\displaystyle\hat\varphi_+(\xi,x_2) &=&\displaystyle \frac{\e^{-i\xi y_1}}{\sqrt{2\pi}}\left\{\frac{\e^{-\gamma_+|x_2-y_2|}}{2\gamma_+}\right. +\left(\frac{\beta_--\beta_+}{\beta_-+\beta_+}\right)\frac{\e^{-\gamma_+|x_2+y_2|}}{2\gamma_+}\medskip\\
&&\displaystyle\left.+\frac{\beta_+\beta_-(k^2_{-}-k^2_+)}{(\beta_-\gamma_++\beta_+\gamma_-)(\beta_-+\beta_+)}\frac{\e^{-\gamma_+ (x_2+y_2)}}{\gamma_+(\gamma_++\gamma_-)}\right\},\bigskip\\
\displaystyle\hat\varphi_-(\xi,x_2) &=&\displaystyle\frac{\e^{-i\xi y_1}}{\sqrt{2\pi}}\left\{\frac{\beta_-}{\beta_-+\beta_+}\frac{\e^{-\gamma_+(y_2-x_2)}}{\gamma_+}+\right. \left. \left(\frac{\beta_-\e^{\gamma_-x_2-\gamma_+y_2}}{\beta_-\gamma_++\beta_+\gamma_-}-\frac{\beta_-}{\beta_-+\beta_+}\frac{\e^{-\gamma_+(y_2-x_2)}}{\gamma_+}\right)\right\},
\end{array}\hspace{-0.7cm}
\end{equation*}
where $\gamma_\pm = \sqrt{\xi^2-k_\pm^2}$.  The determination of
physically admissible branches of the functions $\gamma_\pm(\xi) =
\sqrt{\xi-k_\pm}\sqrt{\xi+k_\pm}$ require selection of branch cuts for
each one of the two associated square root functions. The relevant
branches, which are determined by consideration of Sommerfeld's
radiation condition, are $-3\pi/2 \leq \arg(\xi-k_\pm)<\pi/2$ for
$\sqrt{\xi-k_\pm}$ and $-\pi/2\leq\arg(\xi+k_\pm)<3\pi/2$ for
$\sqrt{\xi+k_\pm}$.  Taking the inverse Fourier
transform~\eqref{eq:fourier_transform} of $\hat\varphi_\pm$ and using
the identity
$$
\int_\R\frac{\e^{-\gamma_\pm|x_2-y_2|}}{4\pi \gamma_\pm}\e^{i\xi(x_1-y_1)}\de \xi = \frac{i}{4}H^{(1)}_0(k_\pm|\ney-\nex|),
$$
we obtain
\begin{eqnarray*}
\varphi_+(\nex) &=&\frac{i}{4} H_0^{(1)}(k_+|\nex-\ney|)+\frac{i}{4}\frac{\beta_--\beta_+}{\beta_-+\beta_+}H_0^{(1)}(k_+|\bar\nex-\ney|)+\Phi^+(\nex,\ney),\\
\varphi_-(\nex) &=&\frac{i}{2}\frac{\beta_-}{\beta_-+\beta_+} H_0^{(1)}(k_+|\nex-\ney|)+\Phi^-(\nex,\ney),
\end{eqnarray*}
where the functions $\Phi_\pm$ are given by
\begin{equation*}
\begin{array}{rcl}
\Phi_+(\nex,\ney) &=& \displaystyle\frac{\beta_+\beta_-(k_-^2-k_+^2)}{\pi (\beta_-+\beta_+)} \displaystyle\int_{0}^{\infty}\frac{\e^{-\gamma_+(x_2+y_2)}\cos(\xi(x_1-y_1))}{\gamma_+(\gamma_-+\gamma_+)(\beta_-\gamma_++\beta_+\gamma_-)}\de \xi,\bigskip\\
\Phi_-(\nex,\ney) &=& \displaystyle\frac{\beta_-}{\pi}\int_{0}^{\infty}\left(\frac{\e^{\gamma_- x_2-\gamma_+ y_2}}{\gamma_+\beta_-+\gamma_-\beta_+}\right. \left.-\frac{\e^{\gamma_+(x_2-y_2)}}{(\beta_++\beta_-)\gamma_+}\right)\cos(\xi(x_1-y_1))\de \xi,
\end{array}
\end{equation*} 
Similarly, the solution of the ODE system for $\ney\in\R^2_-$ is given by 
\begin{equation*}
\begin{array}{lcl}
\displaystyle\hat\varphi_+(\xi,x_2) &=&\displaystyle\frac{\e^{-i\xi y_1}}{\sqrt{2\pi}}\left\{\frac{\beta_+}{\beta_-+\beta_+}\frac{\e^{-\gamma_-(x_2-y_2)}}{\gamma_-}+\right. \left. \left(\frac{\beta_+\e^{-\gamma_+x_2+\gamma_-y_2}}{\beta_-\gamma_++\beta_+\gamma_-}-\frac{\beta_+}{\beta_-+\beta_+}\frac{\e^{-\gamma_-(x_2-y_2)}}{\gamma_-}\right)\right\},\bigskip\\
\displaystyle\hat\varphi_-(\xi,x_2) &=&\displaystyle \frac{\e^{-i\xi y_1}}{\sqrt{2\pi}}\left\{\frac{\e^{-\gamma_-|x_2-y_2|}}{2\gamma_-}\right. +\left(\frac{\beta_+-\beta_-}{\beta_-+\beta_+}\right)\frac{\e^{-\gamma_-|x_2+y_2|}}{2\gamma_-}\medskip\\
&&\displaystyle\left.+\frac{\beta_+\beta_-(k^2_+-k^2_-)\e^{\gamma_- (x_2+y_2)}}{(\beta_-\gamma_++\beta_+\gamma_-)(\beta_-+\beta_+)\gamma_-(\gamma_-+\gamma_+)}\right\}.
\end{array}\hspace{-0.7cm}
\end{equation*}
Taking inverse Fourier transform \eqref{eq:fourier_transform} we now obtain 
\begin{eqnarray*}
\varphi_+(\nex) &=&\frac{i}{2}\frac{\beta_+}{\beta_-+\beta_+} H_0^{(1)}(k_-|\nex-\ney|)+\Psi^-(\nex,\ney),\\
\varphi_-(\nex) &=&\frac{i}{4} H_0^{(1)}(k_-|\nex-\ney|)+\frac{i}{4}\frac{\beta_+-\beta_-}{\beta_-+\beta_+}H_0^{(1)}(k_-|\bar\nex-\ney|)+\Psi^+(\nex,\ney),
\end{eqnarray*}
where the functions $\Psi_\pm$ are given by
\begin{equation*}
 \begin{array}{lcl}
\Psi_+(\nex,\ney) &=& \displaystyle\frac{\beta_+}{\pi}\int_{0}^{\infty}\left(\frac{\e^{-\gamma_+ x_2-\gamma_- y_2}}{\gamma_-\beta_++\gamma_+\beta_-}-\frac{\e^{-\gamma_-(x_2-y_2)}}{(\beta_++\beta_-)\gamma_-}\right)\cos(\xi(x_1-y_1))\de \xi,\bigskip\\
\Psi_-(\nex,\ney) &=& \displaystyle\frac{\beta_+\beta_-(k_+^2-k_-^2)}{\pi (\beta_-+\beta_+)}\int_{0}^{\infty}\frac{\e^{\gamma_-(x_2+y_2)}\cos(\xi(x_1-y_1))}{\gamma_-(\gamma_-+\gamma_+)(\beta_-\gamma_++\beta_+\gamma_-)}\de \xi.
\end{array}
\end{equation*}
The gradient of the Green function is evaluated from the expressions
above by differentiation under the integral sign.

\section{Green function for a two-layer medium: numerical
  computation\label{app:greens_function_comp}} 
In order to evaluate numerically the functions $\Phi_\pm$, $\Psi_\pm$
(Appendix~\ref{app:greens_function}) and their derivatives we use a
contour integration method described in~\cite{PAULUS:2000} together
with the smooth-windowing approach put forth in~\cite{MONRO:2007,brunodelourme}
for evaluation of oscillatory integrals. As an example we consider here the
problem of evaluation of $\Phi_+$, the corresponding problem of
evaluation of $\Phi_-$, $\Psi_\pm$ and derivatives of $\Phi_\pm$ and
$\Psi_\pm$ can be treated similarly. The evaluation of $\Phi_+$
requires integration of the function
$$
\phi(\xi)=\frac{\e^{-\gamma_+(\xi)(x_2+y_2)}\cos(\xi(x_1-y_1))}{\gamma_+(\xi)[\gamma_-(\xi)+\gamma_+(\xi)][\beta_-\gamma_+(\xi)+\beta_+\gamma_-(\xi)]},
$$
which is highly oscillatory for wide ranges of values of the spatial
variables $x$ and $y$, and which is additionally singular at certain
points in the integration domain.

Here we consider the most challenging case in which one or both of the
wavenumbers $k_\pm$ is real, in such a way that $\phi$ has
branch-point singularities at $\xi=k_-\in \R$ and/or $\xi=k_+\in
\R$. Note that significant simplifications occur in the case in which
both media are lossy since, in view of the definition of $\gamma_\pm$,
for lossy media the function $\phi$ is smooth (in fact analytic) on
the whole positive real axis.  Also note that $\phi, \phi_{x_2}$ and
$\phi_{y_2}$ decay exponentially fast as $\xi\rightarrow\infty$ when
$x_2+y_2>0$. However, $\phi$ decays as $|\xi|^{-3}$ and $\phi_{x_2}$
and $\phi_{y_2}$ decay as $|\xi|^{-2}$ as $|\xi|\rightarrow\infty$
when $x_2=y_2=0$.


To proceed with the numerical evaluation of the needed integral of
$\phi$ we write $\int_0^\infty \phi(\xi)\de \xi = I_1+I_2,$ where
$I_1=\int_0^{L_1} \phi(\xi)\de \xi$ and $I_2=\int_{L_1}^\infty
\phi(\xi)\de \xi$, and where $L_1$ is an adequately selected real
number such that $L_1>\max\{\real\, k_-,\real\, k_+\}$. Note that the
branch cuts set forth in Appendix~\ref{app:greens_function} are
vertical rays directly above of the interval $0\leq \xi\leq L_1$; the ray 
end-points $k_\pm$, further, are close to (resp. on) the real $\xi$ axis for
small (resp. vanishing) values of the imaginary parts of
$k_\pm$. Using the Cauchy integral theorem we obtain
\begin{equation}\label{eq:first_integral}
I_1= \int_{C} \phi(z)\de z
=\displaystyle\int_{-1}^1 \phi(\zeta(t))|\zeta'(t)|\de t,
\end{equation}
where  $C$ is a simple curve in the fourth quadrant which is parametrized by $\zeta:[-1,1]\mapsto \C$ satisfying $\zeta(-1)=0$ and $\zeta(1)=L_1$.  

In order to evaluate $I_2$, on the other hand, we utilize the
partition of the unity method introduced
in~\cite{MONRO:2007,brunodelourme}. Hence
\begin{equation}
\begin{array}{rcl}
I_2&=&\displaystyle\int_{L_1}^\infty \phi(\xi)\de \xi \approx\int_{L_1}^{\infty} \phi(\xi)\eta(\xi,cL_2,L_2)\de \xi=\int_{L_1}^{L_2}\phi(\xi)\eta(\xi,cL_2,L_2)\de \xi
\end{array}\label{eq:second_integral}
\end{equation}
where $L_2>L_1$, $L_1/L_2<c<1$  and $\eta$ is the window function defined by
\begin{equation*}
\eta(\xi,\xi_0,\xi_1)\! =\! \left\{
\begin{array}{ccc}
1,&|\xi|\leq \xi_0\\
\!\!\exp\left(\displaystyle\frac{2\e^{-1/u}}{u-1}\right),& \xi_0<|\xi|<\xi_1, u=\frac{|\xi|-\xi_0}{\xi_1-\xi_0},\\
0,&|\xi|>\xi_1.
\end{array}\right.
\end{equation*}
It can be shown that the last integral in (\ref{eq:second_integral})
converges super-algebraically fast to $I_2$ as $L_2$ goes to infinity
\cite{MONRO:2007,brunodelourme}.

Throughout the examples presented in this paper the curve $C$ is the
ellipse $\zeta(t) = \{(L_1+L_1\cos(\pi(t+3)/2))/2+iL_1 \sin
(\pi(t+3)/2)/4, t\in(-1,1)\}$ where $L_1=\real\{k_-+k_+\}$. The last
integral in (\ref{eq:first_integral}) and the last integral in
(\ref{eq:second_integral}) are approximated by using Clenshaw-Curtis
quadrature, which yields rapid convergence for the smooth integrands
under consideration.

\bibliographystyle{abbrv}
\bibliography{References}

%

\end{document}